\documentclass[acmsmall, nonacm]{acmart}

\newcommand{\ACM}{ACM}
\newcommand{\LNCS}{LNCS}
\newcommand{\IEEE}{IEEE}
\newcommand{\FORMAT}{ACM}

\ifx\FORMAT\ACM

\AtBeginDocument{%
  \providecommand\BibTeX{{%
    \normalfont B\kern-0.5em{\scshape i\kern-0.25em b}\kern-0.8em\TeX}}}

\setcopyright{none}

\acmJournal{JACM}
\acmVolume{37}
\acmNumber{4}
\acmMonth{8}

\else

\ifx\FORMAT\LNCS

\usepackage[T1]{fontenc}

\else

\fi

\fi

\definecolor{edgeblue}{RGB}{0, 0, 200}
\definecolor{edgegreen}{RGB}{0, 200, 0}
\definecolor{gptgreen}{RGB}{0, 166, 126}
\definecolor{scholarpurple}{RGB}{169, 1, 251}
\definecolor{bgcode}{rgb}{0.95,0.95,0.95}
\definecolor{githubgreen}{rgb}{0.564, 0.933, 0.564}
\definecolor{orange}{rgb}{1,0.5,0}

\definecolor{codegreen}{rgb}{0,0.6,0}
\definecolor{codegray}{rgb}{0.5,0.5,0.5}
\definecolor{backcolour}{RGB}{245,248,250}
\definecolor{emph}{RGB}{166,88,53}
\definecolor{nightblue}{RGB}{9,49,105}
\definecolor{keywords}{RGB}{207,33,46}
\definecolor{lightpurple}{RGB}{130,81,223}
\definecolor{examplebg}{RGB}{250,243,240}
\definecolor{codemph}{RGB}{150,30,30}

\usepackage{xcolor}

\definecolor{mscolor}{rgb}{0.1,0.1,0.9}

\newcommand{\cmark}{\textcolor{teal}{\ding{51}}}%
\newcommand{\xmark}{\ding{55}}%

\newcommand{\cpplang}{\textsc{C++}\xspace}
\newcommand{\clanguage}{\textsc{C}\xspace}
\newcommand{\rustlang}{\textsc{Rust}\xspace}

\newcommand{\llm}{\textsc{LLM}\xspace}
\newcommand{\llms}{\textsc{LLMs}\xspace}

\newcommand{\sysname}{\textsc{Syzygy}\xspace}
\newcommand{\specminer}{\textsc{SpecMiner}\xspace}
\newcommand{\slicer}{\textsc{Slicer}\xspace}
\newcommand{\codegenerator}{\textsc{CodeGenerator}\xspace}
\newcommand{\argtranslator}{\textsc{ArgTranslator}\xspace}
\newcommand{\eqtester}{\textsc{EqTester}\xspace}

\newcommand{\translateargs}{\textsf{translateArgs}}

\newcommand{\urlparser}{\textsc{UrlParser}\xspace}
\newcommand{\zopfli}{\textsc{Zopfli}\xspace}

\newcommand{\clangseventeen}{\textsc{Clang-17}\xspace}
\newcommand{\llvm}{\textsc{LLVM}\xspace}
\newcommand{\llvmfourteen}{\textsc{LLVM-14}\xspace}

\newcommand{\numzopflitests}{$26$\xspace}
\newcommand{\clinecoverage}{$88$\xspace}
\newcommand{\cbranchcoverage}{$70$\xspace}

\newcommand{\numzopflitestseval}{$1000000$\xspace}
\newcommand{\cevallinecoverage}{$95$\xspace}
\newcommand{\cevalbranchcoverage}{$83$\xspace}

\newcommand{\partialsat}{{\textcolor{orange}{\ding{51}}\textcolor{orange}{\kern-0.6em\Large\ding{55}}}}

\usepackage{blindtext}
\usepackage{xpatch}
\usepackage{amsmath}
\usepackage{amsfonts}
\usepackage{amsthm}
\usepackage{graphicx}
\usepackage[T1]{fontenc}
\usepackage{mathrsfs}
\usepackage{hyperref}
\usepackage{tikz}
\usepackage{tikz-cd}
\usepackage{algorithm}
\usepackage{algorithmic}
\usepackage{multirow}
\usepackage{booktabs}
\usepackage{colortbl}
\usepackage{url}
\usepackage{subfigure}
\usepackage{array}
\usepackage[normalem]{ulem}
\usepackage{xspace}
\usepackage{wrapfig}
\usepackage{tcolorbox}
\usepackage[framemethod=tikz]{mdframed}
\usepackage{enumitem}
\usepackage{balance}
\usepackage{fontawesome}
\usepackage{pifont}
\usepackage{cleveref}
\usepackage{tcolorbox}
\usepackage{tipa}
\usepackage{pifont}

\usepackage{listings}
\usepackage{listings-rust}
\usepackage{xcolor}

\definecolor{mGreen}{rgb}{0,0.6,0}
\definecolor{mGray}{rgb}{0.5,0.5,0.5}
\definecolor{mPurple}{rgb}{0.58,0,0.82}

\lstdefinestyle{CStyle}{
    backgroundcolor=\color{orange!5},
    commentstyle=\color{mGreen},
    keywordstyle=\color{magenta},
    numberstyle=\tiny\color{mGray},
    stringstyle=\color{mPurple},
    basicstyle=\fontencoding{T1}\fontfamily{lmtt}\footnotesize,
    breakatwhitespace=false,         
    breaklines=true,                 
    captionpos=b,                    
    keepspaces=true,
    numbers=left,                  
    numbersep=5pt,                  
    xleftmargin=0.6cm,
    showspaces=false, 
    frame=single, 
    framerule=0.05cm,
    rulecolor=\color{orange!80},
    showstringspaces=false,
    showtabs=false,                  
    tabsize=2,
    language=C
}

\lstdefinestyle{PStyle}{
    backgroundcolor=\color{black!5},
basicstyle=\fontencoding{T1}\fontfamily{lmtt}\footnotesize,
    breakatwhitespace=false,         
    breaklines=true,                 
    captionpos=b,                    
    keepspaces=true,
    numbers=left,                  
    numbersep=5pt,                  
    xleftmargin=0.6cm,
    showspaces=false, 
    frame=single, 
    framerule=0.05cm,
    rulecolor=\color{black!80},
    showstringspaces=false,
    showtabs=false,                  
    tabsize=2,
    language=C
}

\newcommand{\rustc}[1]{\lstinline[language=Rust, style=RStyle, basicstyle=\fontencoding{T1}\fontfamily{lmtt}\small]{#1}}
\newcommand{\rustct}[1]{\text{\lstinline[language=Rust, style=RStyle, basicstyle=\fontencoding{T1}\fontfamily{lmtt}\small]{#1}}}
\newcommand{\cc}[1]{\lstinline[language=C, style=CStyle, basicstyle=\fontencoding{T1}\fontfamily{lmtt}\small]{#1}}
\newcommand{\cct}[1]{\text{\lstinline[language=C, style=CStyle, basicstyle=\fontencoding{T1}\fontfamily{lmtt}\small]{#1}}}

\definecolor{edgeblue}{RGB}{0, 0, 200}
\definecolor{edgegreen}{RGB}{0, 200, 0}
\definecolor{gptgreen}{RGB}{0, 166, 126}
\definecolor{scholarpurple}{RGB}{169, 1, 251}
\definecolor{bgcode}{rgb}{0.95,0.95,0.95}
\definecolor{githubgreen}{rgb}{0.564, 0.933, 0.564}
\definecolor{orange}{rgb}{1,0.5,0}

\definecolor{codegreen}{rgb}{0,0.6,0}
\definecolor{codegray}{rgb}{0.5,0.5,0.5}
\definecolor{backcolour}{RGB}{245,248,250}
\definecolor{emph}{RGB}{166,88,53}
\definecolor{nightblue}{RGB}{9,49,105}
\definecolor{keywords}{RGB}{207,33,46}
\definecolor{lightpurple}{RGB}{130,81,223}
\definecolor{examplebg}{RGB}{250,243,240}
\definecolor{codemph}{RGB}{150,30,30}

\newcommand{\Paragraph}[1]{\smallskip\noindent{\bf #1.}}
\newcommand{\SubParagraph}[1]{\smallskip\noindent{\emph{#1}.}}

\newcommand{\secref}[1]{\S\ref{#1}}

\theoremstyle{definition}

\ifx\FORMAT\ACM

\title{This is a title}

\ifx\FORMAT\ACM

\title[\sysname{}: Dual Code-Test \clanguage{} to (safe) \rustlang{} Translation using \llms{} and Dynamic Analysis]{\sysname{}: Dual Code-Test \clanguage{} to (safe) \rustlang{} Translation using\\ \llms{} and Dynamic Analysis}

\renewcommand{\shortauthors}{Anon}

\author[MS.]{Manish Shetty
}
\authornote{Equal Contribution, order determined using a dice roll.}
\email{manishs@berkeley.edu}
\affiliation{
    \institution{University of California, Berkeley}
    \city{Berkeley}
    \state{CA}
    \country{USA}
}

\author[NJ.]{Naman Jain
}
\authornotemark[1]
\email{naman\_jain@berkeley.edu}
\affiliation{
    \institution{University of California, Berkeley}
    \city{Berkeley}
    \state{CA}
    \country{USA}
}

\author[AG.]{Adwait Godbole
}
\authornotemark[1]
\email{adwait@berkeley.edu}
\affiliation{
    \institution{University of California, Berkeley}
    \city{Berkeley}
    \state{CA}
    \country{USA}
}

\author[SS.]{Sanjit A. Seshia
}
\authornote{Equal Advising, order determined using a coin toss.}
\email{sseshia@berkeley.edu}
\affiliation{
    \institution{University of California, Berkeley}
    \city{Berkeley}
    \state{CA}
    \country{USA}
}

\author[KS.]{Koushik Sen
}
\authornotemark[2]
\email{ksen@berkeley.edu}
\affiliation{
    \institution{University of California, Berkeley}
    \city{Berkeley}
    \state{CA}
    \country{USA}
}

\renewcommand{\shortauthors}{M. Shetty, N. Jain, A. Godbole, S. A. Seshia, K. Sen}

\else

\ifx\FORMAT\LNCS

\else

\ifx\FORMAT\IEEE

\fi

\fi

\fi

\fi

\begin{document}

\ifx\FORMAT\ACM


\ifx\FORMAT\ACM

\begin{abstract}
Despite extensive usage in high-performance, low-level systems programming applications, C is susceptible to vulnerabilities due to manual memory management and unsafe pointer operations.
Rust, a modern systems programming language, offers a compelling alternative. Its unique ownership model and type system ensure memory safety without sacrificing performance. 

In this paper, we present \sysname{}\footnote{\sysname{} (pronounced  \textipa{["sIz.I.dZi]} / \texttt{si.zuh.jee}), derived from ancient Greek, is used to represents the dynamic alignment and conjunction of complementary forces that come together in perfect balance.}, an automated approach to translate C to \textit{safe} Rust.
%
Our technique uses a synergistic combination of LLM-driven code and test generation guided by dynamic-analysis-generated execution information.
This paired translation runs incrementally in a loop over the program in dependency order of the code elements while maintaining per-step correctness.
%
Our approach exposes novel insights on combining the strengths of LLMs and dynamic analysis in the context of scaling and combining code generation with testing.
We apply our approach to successfully translate \zopfli{}, a high-performance compression library with $\sim$3000 lines-of-code and 98 functions generating $\sim$4500 lines-of-code in Rust.
We validate the translation by testing equivalence with the source C program on a broad set of inputs.
To our knowledge, this is the largest automated and test-validated C to \textit{safe} Rust code translation achieved so far.

\begin{center}
{\centering
Project Website: \url{https://syzygy-project.github.io/}
}
\end{center}
\end{abstract}

\else

\ifx\FORMAT\LNCS 

\begin{abstract}
Despite extensive usage in high-performance, low-level systems programming applications, including the Linux kernel, C is susceptible to vulnerabilities, primarily due to manual memory management and unsafe pointer operations.
Rust, a modern systems programming language, offers a compelling alternative. Its unique ownership model and type system ensure memory safety without sacrificing performance. 

We present an automated approach to translate C to \textit{safe} Rust.
Our technique uses a synergistic combination of LLM-driven code generation guided by dynamic-analysis-generated execution information.
Our approach exposes novel insights on scaling, testing, and combining the strengths of LLMs and dynamic analysis.
We apply our approach to successfully translate \texttt{zopfli}, a high-performance compression library with $\sim$3000 LoC and 98 functions.
We validate the translation by testing equivalence with the source C program on a set of inputs.
To our knowledge, this is the largest automated and test-validated C to \textit{safe} Rust code translation achieved so far.

\end{abstract}

\else

\ifx\FORMAT\IEEE

\begin{abstract}
Despite extensive usage in high-performance, low-level systems programming applications, including the Linux kernel, C is susceptible to vulnerabilities, primarily due to manual memory management and unsafe pointer operations.
Rust, a modern systems programming language, offers a compelling alternative. Through its unique ownership model and type system, it ensures memory safety without sacrificing performance. 

We present an automated approach to translate C to \textit{safe} Rust.
Our technique uses a synergistic combination of LLM-driven code generation guided by dynamic-analysis-generated execution information.
Our approach exposes novel insights on scaling, testing, and combining the strengths of LLMs and dynamic analysis.
We apply our approach to successfully translate \texttt{zopfli}, a high-performance compression library with $\sim$3000 LoC and 98 functions.
We validate the translation by testing equivalence with the source C program on a set of inputs.
To our knowledge, this is the largest automated and test-validated C to \textit{safe} Rust code translation achieved so far.
\end{abstract}


\fi 

\fi

\fi

\maketitle

\else

\ifx\FORMAT\LNCS

\maketitle

\else

\ifx\FORMAT\IEEE

\maketitle

\fi
\fi
\fi


\section{Introduction}

C to Rust translation has seen tremendous interest in recent years owing to memory safety vulnerabilities in C \cite{eternalwar} on the one hand and Rust's strong static type and ownership system guaranteeing compile-time eradication of these vulnerabilities on the other.
It is further motivated by the fact that both C and Rust can target similar applications (low-level, performance-critical libraries) and are supported by Clang-based compiler toolchains.
Though desirable, C-Rust translation is challenging: C and (safe) Rust employ different typing systems (strongly typed variables and no raw pointers in Rust) and different variable access rules (arbitrary accesses in C, while strict borrowing rules in Rust), amongst other differences.
Manual migration of even moderately sized codebases requires multiple person-weeks of effort, motivating the need for automatic translation techniques. 
%

There are two main approaches for code translation: rule-based/symbolic and \llm{} (Large Language Model)-based. Rule-based translation approaches often operate on a terse intermediate representation (for achieving full coverage with a limited rule set) and thus often produce uninterpretable target code. Symbolic program synthesis approaches (e.g., \cite{sygus, 10.1145/357084.357090}), on the other hand, often do not scale to multi-function codebases.
\llms{} shine in both these respects: they produce natural/interpretable code and have superior scaling capabilities. 
\llms{} cannot, however, perform precise inference of semantic features of program executions such as aliasing and allocation sizes.
This is especially important in languages with pointer casts and dynamic allocations such as C, where this information is often not syntactically available/interpretable.

We develop \sysname{}, an approach to convert medium-large (multi-file, multi-function) C codebases to equivalent \textit{safe} Rust code automatically. 
As the name suggests, \sysname{}\footnotemark[1] exhibits two kinds of dualities. The first is a synergistic combination of superior generative capabilities of \llms{} with semantic execution information collected by dynamic analyses \cite{10.1145/318774.318944} on the source C codebase.
Secondly, as opposed to using the \llms{}+Dynamic Analysis recipe to perform Rust code-generation \textit{alone}, we also build a test translation approach for reliable equivalence testing.
\begin{figure}
    \centering
    \includegraphics[width=0.9\linewidth]{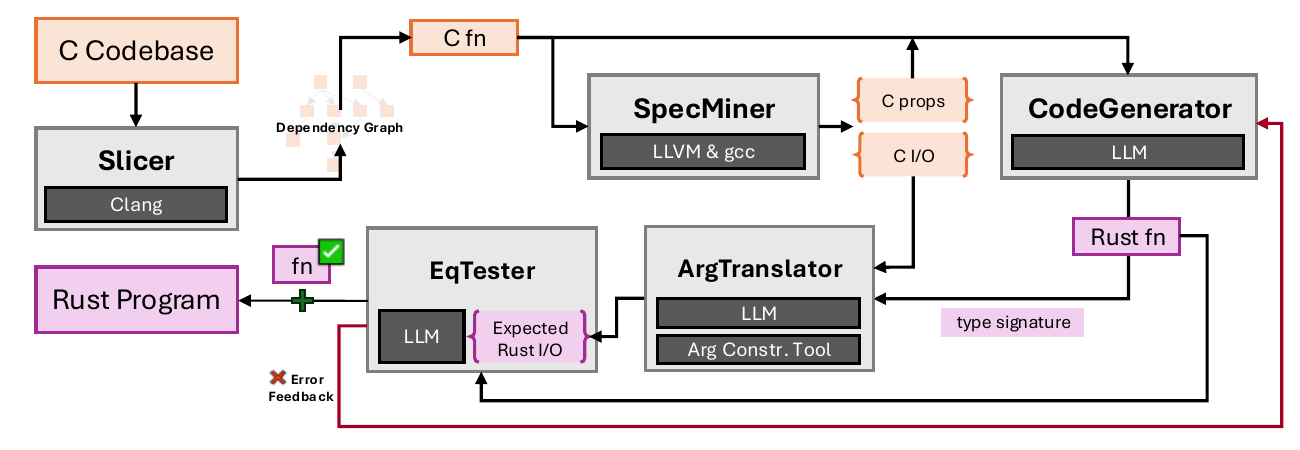}
    \caption{\textbf{Overview of our \sysname{} translation approach}: \slicer decomposes the codebase into translation units, \codegenerator{} performs translation for that unit, and \eqtester{} checks for equivalence of the generated Rust code with the original C code.
    \argtranslator{} maps the \clanguage{} and \rustlang{} function arguments allowing appropriate equivalence checks in \eqtester{}.
    \specminer{}, our dynamic analysis module mines (property and I/O) specifications for intermediate \clanguage{} functions using top-level tests. 
    These specifications later assist our  \llm{}-driven dual code-test translation.
    In particular, our identified properties (nullability, aliasing, types) guide \llm{} to generate appropriate target \rustlang{} signature. 
    Similarly, our collected I/O specifications allow \eqtester{} module to provide correctness assurances on incremental translations generating our \emph{Invariant}.
    }
    \label{fig:overview}
\end{figure}

The four tenets of our approach - \llms{} + Dynamic Analysis and Code + Test Generation work together as follows. 
First, C constructs such as dynamic (heap) allocations of unknown sizes, aliasing between pointers, and pointer type casts (and the lack thereof in Rust) lead to challenging translation scenarios in which execution information (e.g., array sizes, aliasing, and types) must be inferred from the code. Our dynamic analysis extracts this semantic execution information and makes it available to the \llm{}, thus aiding translation.
Second, to efficiently leverage the stochastic generative capabilities of \llms{} for large codebases, we need \textit{tests} to weed out and/or \textit{repair} incorrect candidates. 
While tests for top-level functions are often available/easy to construct (due to documented APIs), reliably testing intermediate functions is challenging.
Here we once again use our LLM+Dynamic Analysis recipe: we (a) use dynamic allocation analysis to mine intermediate function I/O examples from top-level \clanguage{} tests, and then (b) use \llms{} to programmatically map these I/O examples to \rustlang{} to construct intermediate function equivalence tests.

This approach of combining incremental code generation with equivalence test generation (for repair and validation) draws inspiration from test-driven development (TDD) practices improving code design and reliability~\cite{nagappan2008realizing,calais2023test}.
Notably, just as TDD uses tests to drive implementation, our approach uses intermediate equivalence tests
to greedily translate a C codebase while maintaining compatibility with an \textit{invariant}: previously generated Rust code.
%
This allows us to greatly improve the accuracy, and hence scalability, of translation.
We compare our technique with other C-to-Rust approaches in Table \ref{tab:comparison}.


Our approach decomposes (\slicer in Fig. \ref{fig:overview}) the large codebase into individual \textit{translation units} (e.g., functions, macros, type definitions) and then performs Iterative Aligned Translation (IAT) for each unit.
That is, for each unit, we generate its Rust translation that is semantically equivalent and is aligned with respect to the I/O interface (\codegenerator in Fig. \ref{fig:overview}).
We then test (\argtranslator, \eqtester in Fig. \ref{fig:overview}) the generated Rust unit against its C counterpart, perform a multi-turn repair if necessary, and move on to the next translation unit when done.

We implement our approach using LLM query APIs for code/test generation, the \llvm{} toolchain for implementing dynamic analyses, bindgen, and FFI (Foreign Function Interface) for calling C from Rust during equivalence tests.
First, we show that using our proposed Dual Code-Test Translation approach, we can successfully translate \urlparser{}~\cite{urlparser}, not solved by prior approaches~\cite{yang2024vert, li2024translating}.
Next, for our main case study evaluation, we apply the approach/implementation to translate the \zopfli{} \cite{zopfli} high-performance compression library with 98 functions, 10 structs, and over 3000 LoC.
We validate the correctness of our translation by performing equivalence testing on the top-level entry point with one million randomly generated strings used for compression. 
These tests achieve over $\cevallinecoverage\%$ line coverage and $\cevalbranchcoverage\%$ branch coverage\footnote{We identify a large fraction of uncovered code consisting of error states such as early exits and assertions.} on the source C code.
To our knowledge, this is the largest test-validated C to safe Rust translation performed so far.

\begin{table}[]
    \centering
    \begin{tabular}{c|c|c|c}
        \hline
        \textbf{Approach} & \rustc{safe} \textbf{Rust} & \textbf{Validity} & \textbf{Generation Technique} \\ \hline
        C2Rust \cite{c2rustgalois} & {\color{red}\xmark} & Cons, \cmark & Rule-based (using C types in Rust) \\ 
        VERT \cite{yang2024vert} & \partialsat & Test, \partialsat & LLMs with sampling \\
        CROWN \cite{zhang2023ownership} & \partialsat & Cons, \cmark & Boostrap best-effort analysis on C2Rust \\
        Shiraishi, et. al. \cite{shiraishi2024contextaware} & \partialsat & Test, \partialsat & LLMs with sampling \\
        \hline
        \sysname{} (ours) & \cmark & Test, \cmark & LLMs with sampling + test feedback \\ \hline
    \end{tabular}
    \caption{\textbf{Comparison of \sysname{} with other C to Rust translation approaches}. Legend: \partialsat = Partial satisfaction, Cons = follows by construction (generation) approach, Test = test-based equivalence validation.}
    \label{tab:comparison}
\end{table}

\Paragraph{Contributions}
In summary, our contributions are as follows:
\begin{enumerate}
    \item We develop an automated approach for \textbf{test-validated} translation from C to \textbf{safe} Rust. Our approach leverages the synergy between superior generative and search capabilities of \llms{} and semantic execution information collected using Dynamic Analysis.
    \item Our \llms{}+Dynamic Analysis recipe performs \textbf{dual code and test generation}. Particularly, we enable reliable equivalence testing by (a) using dynamic analysis to mine intermediate I/O examples from top-level tests and (b) using \llms{} to translate these examples to Rust.
    \item We demonstrate the approach by translating the \zopfli{} high compression library, which has over 3000 LoC, and validate our translation by performing top-level equivalence tests
\end{enumerate}

\section{Background}
\label{sec:background}

\Paragraph{C constructs: dynamic allocations, pointer casts, and aliasing}
C allows dynamic memory management using \cc{malloc()} and \cc{free()}.
While essential when allocation sizes are not known statically, dynamic (heap) allocations require careful reasoning to prevent memory safety issues.
C provides low-level stack/heap accesses using pointers. Pointers can be manually constructed, modified, and cast to other types (including the \cc{void*}). Further, multiple pointers can \textit{alias}, i.e., point to/reference the same/overlapping memory. 
Dynamic allocations, casts, and aliases challenge translation since they require non-local program reasoning.
Our dynamic analyses (for allocation sizes, types, and aliasing) help combat this.

%

\Paragraph{Memory (Un)safety}
Memory safety requires programs not to encounter (runtime) errors like out-of-bounds accesses (e.g., reading from beyond buffer bounds), use-after-free, and null pointer dereferences.
One can exploit such behaviors to mount attacks such as reading memory to exfiltrate data (e.g., private keys), injecting data/code to perform arbitrary execution, etc.
C is memory-unsafe because it allows direct memory manipulation using pointers, burdening the programmer with the task of avoiding such errors. 
Rust addresses these issues through its strong typing rules, enforcing strict compile-time checks, and an ownership model.
%
This is the motivation for converting C to Rust: We want to transfer functionality from legacy C programs to (safe) Rust programs.

\Paragraph{Rust constructs: ownership model, smart pointers}
Unlike C pointers, references in Rust have usage restrictions: they can be mutable, e.g., \rustc{\&mut T}, or immutable, e.g., \rustc{\&T}.
Rust enforces \textit{borrowing rules}, i.e., you can only have \textit{either} one mutable reference or any number of immutable references to an object at a time, and there cannot be any dangling references.
This rule-based ownership model eliminates memory leaks, dangling pointers, 
and memory safety vulnerabilities.

Borrowing rules make representing certain data structures (e.g., doubly linked lists) challenging. 
To allow more complex ownership and mutability scenarios, Rust provides smart pointers like \rustc{Rc<T>} and \rustc{RefCell<T>}.
\rustc{Rc<T>} is a reference-counted (immutable) smart pointer that allows multiple owners (references) to \rustc{T}-typed data.
\rustc{RefCell<T>} uses runtime checks to enable \textit{interior mutability}, allowing data to be mutated even when there are immutable references.
\rustc{Rc<RefCell<T>>} combined allows multiple owners to mutate shared data, which 
is useful for data structures like doubly-linked lists.
Using smart pointers provides flexibility while maintaining safety, albeit with some runtime overhead and the potential for panics if misused.


\section{Problem Formulation}

We aim to perform source-level translation from a C codebase to an equivalent \rustc{safe} Rust program. 
We formulate this as a program-synthesis task: \textit{given as specification a C program, synthesize a safe Rust implementation that satisfies it by preserving its functional behavior}.
\begin{align*}
\text{Given:} \quad & \text{a C program } P_C \in \mathcal{C}, \quad \text{ a set of test inputs } T \subset \mathcal{I}, \\
\text{Find:} \quad &  \text{a Rust program } P_R \in \mathcal{R}_{\text{safe}}, \\
\text{Such that:} \quad & \forall t \in T.~ P_C(t) \simeq P_R(t).
\end{align*}
We check conditions for a satisfactory synthesis using the notion of ``observational equivalence''~\cite{obsequiv}, i.e., the generated Rust program produces equivalent outputs to the C program on a set of test inputs $T$ in the space of inputs $I$.
In addition to generating equivalent Rust code, our main motivation for C to Rust translation is eliminating memory safety vulnerabilities. To do so, we disallow any \rustc{unsafe} blocks in the generated Rust program.
Notably, the two main requirements described above (equivalence and memory safety) may conflict: e.g., some input tests may exercise unsafe behaviour during execution (not compilation). As a consequence of enforcing the \rustc{safe} fragment of Rust, we allow undefined behaviors (possibly runtime panics) in such cases. Next, we discuss details about the scope of the input C and output Rust programs we consider:

\Paragraph{Input C Program} We restrict the input C program to the following conditions:
\begin{enumerate}
    \item Acyclic data structures: data structures with pointer cycles require special care in Rust to avoid memory leaks (e.g., \rustc{Weak} pointers to free disconnected memory cycles).
    \item No multithreading: requires \rustc{Arc}-wrapped references to thread-safely perform borrowing.
    \item No type punning: performing raw memory accesses on untyped or multi-typed memory regions would hinder the best effort type analysis that our approach aims to perform.
\end{enumerate}
These are soft restrictions on our Rust code generation pipeline since we use an \llm{}'s capability to perform the task.
However, our current testing and analysis infrastructure is limited to these cases.

\Paragraph{Output Rust Program}
We forbid all \rustc{unsafe} behaviors in the generated Rust program, irrespective of whether the unsafety is \textit{exploitable} or \textit{unexploitable}.
In general, C programs may purposefully leverage unexploitable unsafety for performance or low-level control (e.g., \cite{rustforlinux}). 
However, distinguishing between exploitable and unexploitable is extremely challenging; hence, we take the conservative approach here. We discuss the challenges and implications of this in \secref{subsec:challenges}.

\section{Approach}
\label{sec:approach}
Our high-level approach, \sysname{} is illustrated in Figure~\ref{fig:overview} with an overview provided in \secref{subsec:appoverview}.
At a high level, our approach follows a dual-translation approach--we incrementally translate both code and tests and use execution filtering to ensure validity.
This dual translation is guided by synergistically combining \llms{} with dynamic analysis.

\begin{figure}[t]
    \centering
    \includegraphics[width=0.95\linewidth]{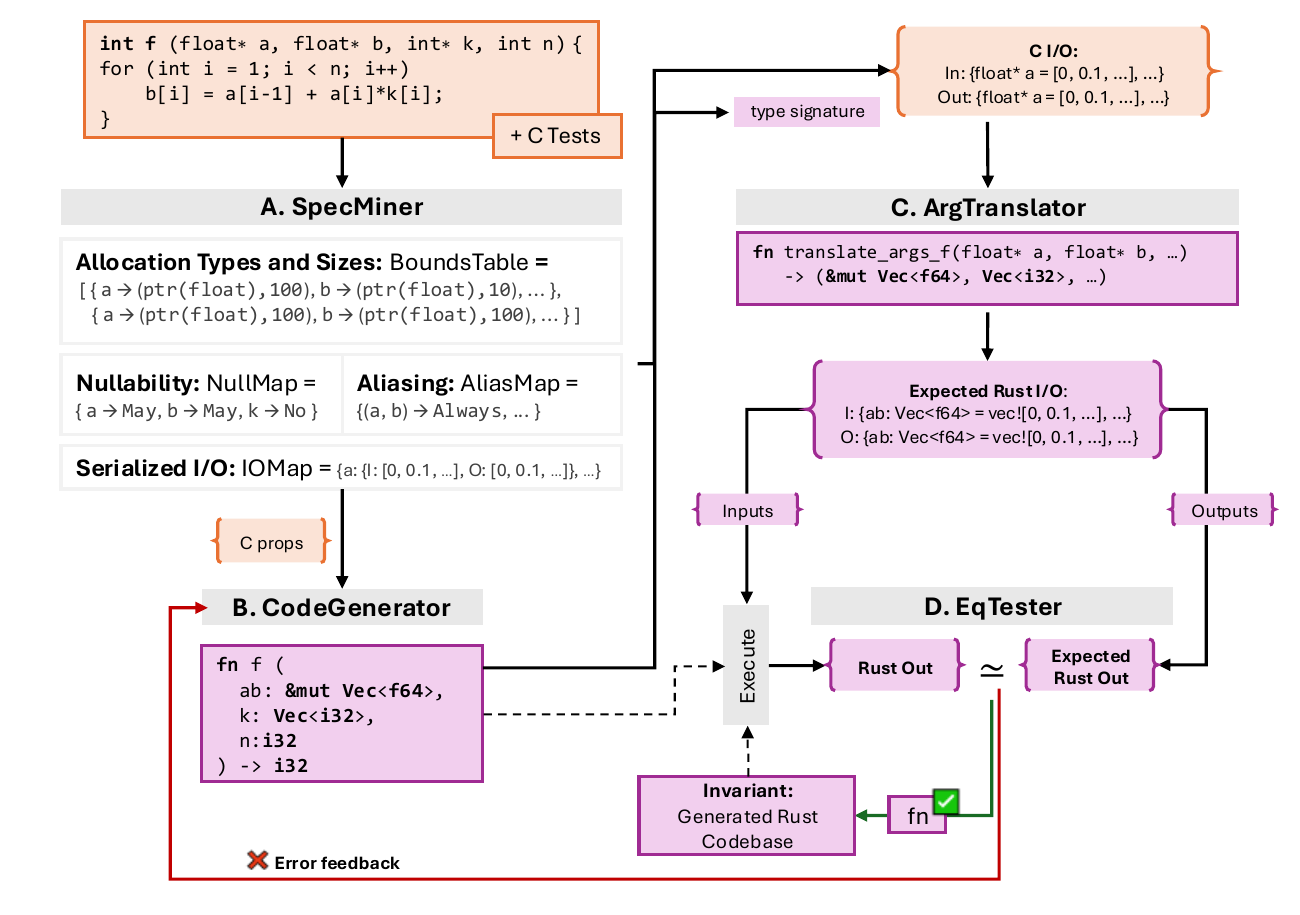}
    \caption{\textbf{Translation Pipeline for a C function \cc{f}}: Given a C function and tests for the C codebase, first, the \specminer{} uses dynamic analysis to mine input-output (I/O) and property specifications for the translation. These properties (e.g., nullability and aliasing) guide a \codegenerator{} to generate a candidate Rust translation of \cc{f}. Given the translated signature and the collected C I/O, the \argtranslator{} generates an \translateargs{} function that constructs aligned Rust inputs and expected outputs. Finally, the \eqtester{} generates an equivalence test that executes the candidate translation, along with the previously generated Rust code, and checks whether the outputs match the expected values.
    If the equivalence checks pass, the translated function is committed to the Rust codebase; otherwise, error feedback is provided for multi-round repair.
    }
    \label{fig:pipeline}
\end{figure}

\subsection{Overview}
\label{subsec:appoverview}

\subsubsection{Incremental Aligned Translation}
Our approach performs translation at the granularity of \textit{translation units} such as functions, structs, \cc{typedef}s, and macros.
That is, we decompose the C codebase into granular units (\slicer in Figure \ref{fig:overview}) and loop over them, translating each incrementally.
We call this incremental aligned translation because the translated Rust code aligns with the source C code primarily in terms of the function signatures and behavior, as well as the function call graph, structs, and type definitions.
Incremental translation, in this way, decomposes code generation and enables the \llm{} to focus on individual functions, thus improving accuracy.

\subsubsection{Semantic Type Alignment \& Dynamic Analysis}
\label{subsec:semtypealign}

Translation between imperative typed languages can be conceptually thought of as a combination of two operations: (1) converting/lifting types of variables and (2) translating expressions and statements on these variables.
%
%
%
The key challenge here arises due to pointers in C. This is because \textit{pointers hide information regarding the underlying object structure and how it's used.}
(Safe) Rust does not allow C-like pointers, thus translation needs to recover this hidden information.
In addition to all of these void pointers in C further obfuscate information by also hiding the type of the underlying object.
We now discuss this challenge further through the following table.
\begin{table}[h]
    \centering
    \begin{tabular}{c | c | c}
        \hline
        Scenario & C type & Possible Rust type
        \\ \hline 
        Array pointer & \cct{(T *)} & \rustct{Vec<T>}, \rustct{[T; N]} \\
        Singleton pointer immutable & \cct{(T *)} & \rustct{& T} \\
        Singleton pointer mutable & \cct{(T *)} & \rustct{&mut T}, \rustct{RefCell<T>}, \rustct{Rc<RefCell<T>>} \\ \hline
        Standard objects (e.g., strings) & \cct{(char *)}& \rustct{crate::string}, \rustct{Vec<u8>}, \rustct{[u8; N]}  \\ \hline
    \end{tabular}
\end{table}
%
\paragraph{Pointers hide object structure}
A C pointer can be used for a singleton object and arrays. The target corresponding Rust type may vary between these cases.
Further, C does not have standard library objects (e.g., strings); rather, these are just represented as \cc{char}/\cc{uchar} arrays. 

\paragraph{Pointers hide usage information}
For example, if argument \cc{a} in function \cc{f} (Fig. \ref{fig:example}) aliased with a pointer from an outer scope, the Rust version of \cc{f} would have to make sure that the shared object (and not a copy) was modified.
Although this is easier in C due to full control over pointers reads/writes, Rust imposes strict borrowing and mutability rules that must be followed: (\rustc{&}, \rustc{&mut}, \rustc{RefCell<T>}, \rustc{Rc<RefCell<T>>}).
In addition to type universality, C pointers enable arbitrary reads and writes, which is not the case in Rust. In Rust, we must specify mutability and nullability and carefully manage aliases between pointers.

\paragraph{Dynamic Analysis} While LLM-driven code generation is powerful, we observe that such aspects of programs are difficult to reason about purely syntactically.
Thus, we develop dynamic analyses (\specminer{}, \secref{subsec:dynamicspec}) to collect properties of function arguments such as aliasing, nullability, and allocation sizes.
We use these to guide the \llm{} during code and test generation.
We use instrumented execution to collect this since static approaches require complex/infeasible program analysis.

\subsubsection{Sampling for Reliable Code Translation}
For each C translation unit, an \llm{} generates corresponding Rust code (\codegenerator, \secref{subsec:codegen}). However, \llms{} can generate incorrect code, making their use unreliable.
Prior work has shown that scaling the number of attempts \llms{} take at problems can help increase performance~\cite{llmmonkeys, alphacode}. Following this,  we use an intuitive approach by \textit{sampling} multiple Rust code generations and \textit{testing} them to filter valid generations. By relying on tests, we transfer the soundness of our approach to the reliability of our tests.

\subsubsection{Reliable Testing by I/O Translation}
However, tests take a black-box approach to verifying code and can have issues such as incompleteness, low code coverage, or even incorrectness. Simply generating tests using \llms{} for arbitrary Rust code does not work due to complexities with invalid types, inter-argument constraints like aliasing, and incorrect assertions on expected outputs.
We address this by observing that C codebases contain tests, particularly for top-level entry points (e.g., main). Invoking the entry point results in calls to internal functions whose inputs and outputs can be captured. 
Consequently, we make testing reliable using \llms{} to programmatically translate these C test inputs to Rust rather than generating from scratch (\argtranslator{}, \secref{subsubsec:argumenttranslation}). 
Then, we validate the translated Rust unit using \llm{} generated value-based equivalence tests (\eqtester, \secref{subsec:testgen}).
These intermediate equivalence tests provide two levels of guarantee about the translated Rust unit: 
(a) it behaves similarly to the source C unit, and 
(b) it maintains compatibility with all previously generated Rust code (\textit{Invariant} in Fig. \ref{fig:pipeline}).
If the test fails, we perform multi-round repair (\secref{subsec:irs}); otherwise, we commit the translated unit to the Rust codebase.


\subsection{Dynamic Specification Mining}
\label{subsec:dynamicspec}
We use dynamic analysis to assist the \llm{} when performing codegen-testgen-repair in our approach.
This analysis produces different kinds of \textit{specifications} (information) about the execution of the C program.
For each C function $f_C$, we gather two kinds of information. 
Firstly, collect \textit{specifications/properties} of the arguments to $f_C$, such as types, bounds, nullability, and aliasing (\secref{subsubsec:c-prop-spec}).
Secondly, we obtain \textit{input-output} examples for $f_C$ (\secref{subsubsec:c-io-spec}). 

We use the mined specifications in two ways. First, this information guides LLM-driven code generation by providing hints distilled into the \llm{}'s prompt (\secref{subsec:codegen}).
Secondly, collected I/O samples are used in equivalence test generation with \llm{} generated tests
(\secref{subsec:testgen}).
Below, we describe the kinds of specifications we collect using the function \cc{f} in Fig. \ref{fig:example} as an example.

\begin{figure}
    \centering
\begin{lstlisting}[style=CStyle, xleftmargin=0cm, basicstyle=\fontencoding{T1}\fontfamily{lmtt}\scriptsize]
float kernel1(float a, float b, int k) { return b - a*k; }
  
void f(float* arr, float* brr, int* krr, int num, float (*kernel)(float, float, int)) { 
    for (int i = 1; i < num; i++)
        brr[i] = (*kernel)(arr[i-1], arr[i], krr[i]);
}
  
float* caller1() {
    int sizea = 100, sizeb = 10;
    float* arr = malloc(sizea*sizeof(float));
    float* brr = malloc(sizeb*sizeof(float));
    int* krr = malloc(sizea*sizeof(int));
    // populate arr, brr, krr ...
    f(arr, brr, krr, min(sizea, sizeb), kernel1);
    return brr;
}

float* caller2() {
    int size = 100;
    float* arr = malloc(size*sizeof(float));
    int* krr = malloc(size*sizeof(int));
    // populate arr, krr ...
    f(arr, arr, krr, size, kernel1);
    return arr;
}

int main() {
    float* res;
    if (...) { 
        res = caller1(); 
    } else { 
        res = caller2(); 
    }
}
\end{lstlisting}
    \caption{\textbf{Dynamic Specification Mining}: An example C code snippet which features dynamic allocations (in \cct{caller1} and \cct{caller2}), function pointers (in \cct{f}), aliasing between arguments (in the \cct{main} -- \cct{caller2} -- \cct{f} call chain). While this information can greatly assist \codegenerator and \eqtester, inferring it statically is challenging. We use a combination of dynamic analyses (see \secref{subsubsec:c-prop-spec}).}
    \label{fig:example}
\end{figure}

\newcommand{\typeinfo}{\mathsf{TypeInfo}}
\newcommand{\sizeinfo}{\mathsf{SizeInfo}}
\newcommand{\nullinfo}{\mathsf{NullInfo}}
\newcommand{\aliasinfo}{\mathsf{AliasInfo}}

\subsubsection{Mining Specifications from C Executions}
\label{subsubsec:c-prop-spec}

Our approach relies on robust analyses to collect information about function arguments. 
This information is consumed by code and test generation (see \secref{subsubsec:c-io-spec}).
We use dynamic analysis to collect this information (\textit{per execution}) since static approaches require complex program analysis (and may even be infeasible).

Our approach operates by (a) maintaining a runtime that tracks variables and their properties and (b) instrumenting the C codebase (using the LLVM compiler toolchain~\cite{llvm}) with handlers that interact with the runtime.
Then, simply compiling and executing the instrumented codebase allows us to collect information about the types, bounds, nullability, and aliasing for function input and outputs.
Depending on the property, we either aggregate or maintain context-sensitive information \textit{per execution} (e.g., nullability is aggregated and bounds are not)
Our approach is general and extensible also to collect other kinds of dynamic information which may be useful in code/test generation.
We now discuss the particular specifications we collect along with how they are used in the pipeline, using the code in Fig. \ref{fig:example} as a running example.

\Paragraph{Types}
\phantomsection{}
\label{para:types}
We first map inputs and outputs to their types. These can be recovered from the \llvm{} IR types with an instrumentation pass. 
In particular, we track the following types:
\begin{align*}    
    \text{Types}~(\cct{T}) = \text{Primitives}~~ (\cct{int}, \cct{uint}, \cct{char}, \cdots) ~|~ \text{Structs} ~|~ \text{Data pointers to } \cct{T} ~|~ \text{Function pointers}
\end{align*}
We store a mapping from function arguments to their types in our instrumented runtime. For example, in function \cc{f} (Fig. \ref{fig:example}), we would extract the following type mapping:
\begin{align*}
    \typeinfo = \{&\cct{arr} \mapsto \texttt{\small pointer(float)}, \cct{brr} \mapsto \texttt{\small pointer(float)}, \\
    &\cct{krr} \mapsto \texttt{\small pointer(int)}, \cct{num} \mapsto \texttt{\small int}, \cct{kernel} \mapsto \texttt{\small function}\}
\end{align*}

Our type inference process manages several challenging cases. When encountering a \cc{malloc} call, the initial allocation returns a void pointer (\cc{void*}) representing untyped memory. Our \llvm{} pass tracks subsequent \cc{bitcast} instructions that convert this raw memory to typed pointers. Consider the example in Figure \ref{fig:example}, where \cc{caller2} allocates arrays of floats and integers.
For each allocation, we observe a bitcast from \cc{i8*} (\llvm{}'s representation of \cc{void*}) to the target pointer type. By tracking these bitcast operations, we recover the intended type of the allocated memory.
We employ a conservative approach for void pointers, mapping them to a base character type while maintaining pointer semantics. This strategy extends to function pointers, where we preserve the complete function signature during analysis.
Finally, we treat arrays and pointers equivalently: singleton arrays are pointers (with size 1), and arrays are pointers with a larger size. The allocation size information captures this difference (see below).

\Paragraph{Allocation sizes} 
\phantomsection{}
\label{para:allocsize}
Dynamic allocations lead to a major challenge with C-based translation.
Our dynamic analysis tracks allocation size by maintaining a bounds table at runtime.
%

The bounds table maintains comprehensive allocation information in an interval tree data structure. Each allocation record contains metadata, memory bounds, and inner element sizes for nested types. This information is collected through two mechanisms: compiler-inserted instrumentation for stack variables and allocation function interception for heap memory. We maintain type-aware size tracking for both allocation types, accounting for structure fields and array dimensions.
Allocation size information is later used when collecting serialized I/O examples for functions.
For our example function \cc{f}, in the \cc{caller1} path, we would record the following bounds information:
\begin{align*}
    \sizeinfo = \{
        \text{\cc{arr}} \mapsto 100, \text{\cc{brr}} \mapsto 10, 
        \cdots 
    \}
\end{align*}

\Paragraph{Nullability} 
\phantomsection{}
\label{para:nullability}
We also determine nullability by dynamically analysing argument values in various test (execution) contexts.
Given a function with pointer arguments, e.g., \cc{f(int* a)}, we can determine whether \cc{a} is ever \cc{NULL}. If so, we make this information available to the \llm{}.
The same argument can be \cc{NULL} in some executions and non-\cc{NULL} in others. We build a map from arguments to booleans, identifying whether the argument is ever \cc{NULL} (in some execution).
This map directs whether \rustc{Option} type must be applied on an argument.
For example, for function \cc{f}, if the \cc{arr} argument is never \cc{NULL}, then:
\begin{align*}
    \nullinfo = \{\text{\cc{arr}} \mapsto \text{\cc{false}}\}
\end{align*}

\Paragraph{Aliasing} 
\phantomsection{}
\label{para:aliasing}
Additionally, we use dynamic analysis to maintain a map that tracks aliasing between pointer arguments to a function.
This is a map from pairs of function arguments to one of three possible aliasing occurrences: $\{\texttt{Always}, \texttt{Some}, 
\texttt{Never}\}$.
Aliasing between arguments leads to varied signature choices on the Rust side:
\begin{enumerate}
    \item \rustc{&mut T} (collapsed mutable reference): if the arguments \textit{always alias} then we can collapse the two into a single Rust mutable reference argument.
    \item \rustc{Rc<RefCell<T>>}: if arguments \texttt{Some}times alias, then, we might need to maintain full flexibility and use smart pointers (\rustc{Rc<RefCell<T>>}) to allow aliasing in \rustlang{}.
\end{enumerate}
For example, for the function \cc{f}, we end up with the following map, since arguments \cc{arr} and \cc{brr} alias via the \cc{caller2} path, but they don't via the \cc{caller1} path.
\begin{align*}
    \aliasinfo = \{(\text{\cc{arr}}, \text{\cc{brr}}) \mapsto \texttt{Some}, (\text{\cc{arr}}, \text{\cc{krr}}) \mapsto \texttt{Never}, (\text{\cc{brr}}, \text{\cc{krr}}) \mapsto \texttt{Never}\}
\end{align*}

\subsubsection{C I/O Specification}
\label{subsubsec:c-io-spec}

Recall that we validate the correctness of each translated function through test-based equivalence
(against the corresponding C function).
This requires input-output test examples for each function $f_C$ in the codebase.

\Paragraph{Valid inputs}
Determining valid inputs to every function $f_C$ in the codebase is challenging.
For example, functions can have complex inter-argument pre-conditions (e.g., two array arguments must have the same length), pointer arguments may have to be non-\texttt{NULL} or alias each other in specific ways.
Inferring these pre-conditions (towards generating valid inputs) is hard and requires expensive program analysis.
\llms{} similarly struggle to generate valid inputs satisfying such pre-conditions.
We address this by observing that calling the top-level (e.g., main) entry point results in calls to internal functions with valid inputs.
Further, the top-level entry point often has much simpler input constraints and/or is better documented.
Thus, we invoke (fuzz) the top-level function on multiple inputs and collect function call arguments for each resulting execution.

\newcommand{\ppre}{\mathsf{I}}
\newcommand{\ppost}{\mathsf{O}}

\Paragraph{Internal function input capture} To capture I/O examples for function $f_C$, we instrument the entry and exit points of $f_C$ with handlers that dump serialized input and output values.
We serialize values of non-pointer objects (e.g., primitive types) directly. For pointers, we dereference the argument (possibly multiple times) to store the internal object.
At the function's exit point, we dump both the output and the input arguments to ensure we capture side effects.
This gives us an input-output test example: $(\ppre_\clanguage, \ppost_\clanguage)$.
We use these C I/O examples to generate equivalent Rust test inputs ($\ppre_\rustlang$) and to check equivalence (see \secref{subsec:testgen}).

\begin{figure}[t]
    \centering
\hspace{-0.5cm}
\begin{minipage}{0.45\textwidth}
\begin{lstlisting}[style=CStyle, xleftmargin=0cm, basicstyle=\fontencoding{T1}\fontfamily{lmtt}\scriptsize]
typedef struct S { 
  int flag; float * arr; 
} S;

void s_faxpy (S * s1, S * s2) {
  // f from Fig. 3
  int * krr = malloc(...);
  // ... initialize krr
  f(s1->arr, s2->arr, krr, n, kernel1);
}
\end{lstlisting}
\end{minipage}
\hspace{0.8cm}
\begin{minipage}{0.16\textwidth}
\begin{lstlisting}[style=RStyle, xleftmargin=0cm, basicstyle=\fontencoding{T1}\fontfamily{lmtt}\scriptsize]
struct S {
  flag: i32, 
  arr: [f32; N]
}
\end{lstlisting}
\begin{lstlisting}[style=RStyle, xleftmargin=0cm, basicstyle=\fontencoding{T1}\fontfamily{lmtt}\scriptsize]
struct S {
  flag: i32, 
  arr: Vec<f32>
}
\end{lstlisting}
\end{minipage}
\hspace{0.5cm}
\begin{minipage}{0.1\textwidth}
\begin{tikzpicture}

\node[draw] (A) at (2,0) {\cct{struct S}};
\node[draw, rectangle] (F) at (3.5,0) {\cct{f}};
\node[draw, rectangle] (SDAXPY) at (3,1) {\cct{s_faxpy}};
\node[] (ts) at (3,1.75) {$\cdots$};
\node[] (bs) at (3.5,-0.75) {$\cdots$};

\draw[->] (A) -- (SDAXPY);
\draw[->] (F) -- (SDAXPY);
\draw[->] (SDAXPY) -- (ts);
\draw[->] (bs) -- (F);

\end{tikzpicture}
\end{minipage}
\caption{Struct field choices induce function signatures higher in the call chain. Left: C struct \cc{S} and a function \cc{s_faxpy} dependent on the struct. Middle: two choices for the \rustc{arr} field in the struct. Right: a dependency graph showing relation of \cc{S} with other functions.}
\label{fig:structs}
\end{figure}

\subsection{Rust Code Generation}
\label{subsec:codegen}


\subsubsection{High-level strategy: slicing and greedy translation}

As discussed in \secref{subsec:appoverview}, our approach performs incremental aligned translation: we use the program dependency graph to
slice the \clanguage codebase into individual \textit{translation units} and then translate them starting from the leaves (units with no dependencies) upwards in isolation.
Translation units are top-level statements in the C program: functions, struct definitions, macros, \cc{typedef}s.
Operating on individual units at a time, rather than entire files or the full codebase, ensures focused translation and increases accuracy.

\Paragraph{Manual struct definitions} 
C structs often form leaf nodes in the dependency graph (the exception being if they use a macro definition).
The choice of struct fields, however, induces type/code generation decisions higher up in the dependency graph.
Figure \ref{fig:structs} illustrates an example where the struct \cc{S} is used by the function \cc{s_faxpy} (float a-x-plus-y), which in turn calls function \cc{f} (from Fig. \ref{fig:example}).
The choice of the Rust struct \rustc{arr} fields (middle Fig. \ref{fig:structs}) induces decisions elsewhere in the dependency graph. Depending on whether it is an array or a \rustc{Vec}, either \rustc{f} must take array arguments, or \rustc{s_faxpy} must perform the appropriate type conversions. 

These constraints may generally propagate through large parts of the function dependency graph.
In our incremental aligned translation approach, incompatibility may be detected later on in the translation process, which may require backtracking (see also \hyperref[disc:deporder]{Discussion}).
To avoid this, our current approach requires manual specification of struct definitions.
%
This is not a challenge for functions since we can infer function-level usage constraints from I/O examples and tests. 

\subsubsection{Per-unit translation}
For each translation unit, we use \llms{} to generate candidate \rustlang{} translations. 
We follow the following two principles to identify valid compiling translations.


\Paragraph{Using information from dynamic analysis}
We collect property specifications using \specminer{} module providing type, nullability, and aliasing information about the function arguments.
These properties constrain the signature for \rustlang{} translation and we provide them to \llm{} as additional context.
For example, in Figure~\ref{fig:nullability-codegen}, our nullability map allows \llm{} to infer \textit{non-intuitive} properties and correctly generate the correct code. 
Aliasing information similarly directs the function signature in terms of using smart pointer (\rustc{Rc<RefCell<RT>>}) or leveraging argument collapsing strategies.
\begin{figure}[h!]
    \centering
\hspace{-0.5cm}
\begin{minipage}{0.45\textwidth}
\begin{lstlisting}[style=CStyle, xleftmargin=0cm, basicstyle=\fontencoding{T1}\fontfamily{lmtt}\scriptsize]
static size_t EncodeTree (
    const unsigned *ll_lengths,
    const unsigned *d_lengths,
    int use_16, int use_17, int use_18,
    unsigned char *bp,
    unsigned char **out, 
    size_t *outsize
)
\end{lstlisting}
\end{minipage}
\hspace{0.8cm}
\begin{minipage}{0.4\textwidth}
\begin{lstlisting}[style=RStyle, xleftmargin=0cm, basicstyle=\fontencoding{T1}\fontfamily{lmtt}\scriptsize]
pub fn encode_tree( 
    ll_lengths: &[u32],
    d_lengths: &[u32],
    use_16: i32, use_17: i32, use_18: i32,
    bp: Option<&mut u8>,
    out: Option<&mut Vec<u8>>,
    outsize: Option<&mut usize>,
) -> usize {
\end{lstlisting}
\end{minipage}
\caption{
Nullability information guides \rustlang{} function signature. 
The arguments \cc{bp}, \cc{out}, \cc{outsize} of \cc{EncodeTree}  function can be \cc{NULL}.
However, the \llm{} cannot infer these properties from function context and generates an incorrect signature without \rustc{Option}, failing equivalence.
Property specifications inferred using our dynamic analysis allow \llm{} to generate the correct signature on the right (using \rustc{Option}).
}
\label{fig:nullability-codegen}
\end{figure}

\Paragraph{Sampling-driven search}
We use \llms{} to generate multiple translation candidates for a given translation unit. 
In the \codegenerator{} module, we only focus on achieving compiling solutions and defer execution correctness to the \eqtester{} module.
\subsection{Rust Test Generation}
\label{subsec:testgen}

Given a newly generated Rust function $f_R$, the goal is to check for equivalence between $f_R$ and the source C function $f_C$.
We check equivalence using tests created in two steps. First, we translate input C arguments to \textit{equivalent} Rust arguments (\secref{subsubsec:argumenttranslation}). Then, functions $f_C$ and $f_R$ are invoked on these inputs, and their outputs are compared for equivalence (\secref{subsubsec:eqtestgen}).

\begin{figure}[!h]
    \centering
\begin{minipage}{0.25\textwidth}
\begin{lstlisting}[style=CStyle, xleftmargin=0cm, basicstyle=\fontencoding{T1}\fontfamily{lmtt}\scriptsize]
void square (int* a) {
 // In-place square
 *a = (*a) * (*a);
}
\end{lstlisting}
\end{minipage}
\hspace{0.5cm}
\begin{minipage}{0.36\textwidth}
\begin{lstlisting}[style=CStyle, xleftmargin=0cm, basicstyle=\fontencoding{T1}\fontfamily{lmtt}\scriptsize]
int sum (int* a) {
 int s = 0; int i = 0;
 for (; i < SIZE; i++) s += i;
 return s;
}
\end{lstlisting}
\end{minipage}
\hspace{0.5cm}
\begin{minipage}{0.28\textwidth}
\begin{lstlisting}[style=CStyle, xleftmargin=0cm, basicstyle=\fontencoding{T1}\fontfamily{lmtt}\scriptsize]
void caps (char* s) {
 int i = 0;
 for (; i < SIZE; i++)
  s[i] = toupper(s[i]);
}
\end{lstlisting}
\end{minipage}

\begin{minipage}{0.25\textwidth}
\begin{lstlisting}[style=RStyle, xleftmargin=0cm, basicstyle=\fontencoding{T1}\fontfamily{lmtt}\scriptsize]
fn square (a: &mut u32) {
 // In-place square using mutable reference
 *a = (*a) * (*a);
}
\end{lstlisting}
\end{minipage}
\hspace{0.5cm}
\begin{minipage}{0.36\textwidth}
\begin{lstlisting}[style=RStyle, xleftmargin=0cm, basicstyle=\fontencoding{T1}\fontfamily{lmtt}\scriptsize]
fn sum (a: Vec<u32>) -> u32 {
 let mut s = 0;
 for i in a { s += i; }
 s
}
\end{lstlisting}
\end{minipage}
\hspace{0.5cm}
\begin{minipage}{0.28\textwidth}
\begin{lstlisting}[style=RStyle, xleftmargin=0cm, basicstyle=\fontencoding{T1}\fontfamily{lmtt}\scriptsize]
fn caps(s: &mut String) {
 *s = s.to_uppercase();
}
\end{lstlisting}
\end{minipage}
\caption{Challenges with mapping function arguments from C to Rust: both \cc{square} and \cc{sum} C functions have \cc{int*} arguments while their Rust arguments are different since the \cc{a} argument in \cc{sum} points to a single \cc{int} while that in \cc{sum} points to an array of \cc{int}s. The translation is further complicated by cases where C arrays may map to non-arrays in Rust, e.g., in \cc{caps}.}
    \label{fig:signature}
\end{figure}

\subsubsection{C to Rust I/O Translation}
\label{subsubsec:argumenttranslation}

Consider an input-output example ($\ppre_\clanguage$, $\ppost_\clanguage$) for a C function $f_C$ gathered during dynamic specification mining (\secref{subsubsec:c-io-spec}).
We now want to compare the behavior of $f_R$ against $f_C$ for this example.
This requires translating the C argument objects $\ppre_\clanguage$ and $\ppost_\clanguage$ into equivalent Rust objects $\ppre_\rustlang$ and $\ppost_\rustlang$.
The challenge is that \textit{argument translation must adapt to the context-dependent signature of $f_R$ that is chosen by code generation} (\secref{subsec:codegen}). 

For example, in Fig. \ref{fig:signature}, while all three C functions \cc{square}, \cc{sum}, and \cc{caps} take a single pointer argument, they reference different objects: 
a singleton integer, an array of integers, and a string, respectively.
Consequently, the Rust translations of these functions have different signatures. Particularly, as the \rustc{String} example shows, even hardcoding C array pointer to Rust array/\rustc{Vec} mappings may be inadequate.
As such, it is very hard to cover all possible cases using rule-based mappings.
In light of this, we rely on using the \llm{} itself to generate a translation mapping function, \translateargs{}, that maps C objects to Rust objects (both for inputs and expected outputs):
\begin{align*}
    \translateargs{}(\ppre_\clanguage) \mapsto \ppre_\rustlang \qquad \translateargs{}(\ppost_\clanguage) \mapsto (\text{expected})~\ppost_\rustlang 
\end{align*}

\emph{Programmatically translating I/O by generating such translation mapping alleviates the stochastic nature of \llms{}, providing more reliability when translating many tests.
}


\subsubsection{Argument Construction LLM Tool}
\label{subsubsec:translationIR}

Certain aspects of the argument translation, however, require broader program reasoning.
Examples include aliasing between pointer arguments and allocation sizes (e.g. when size is not explicitly available as another argument).
In the \cc{sum} example (Fig. \ref{fig:signature}), constructing 
a \rustc{Vec} from the array pointer would require size information.
Our dynamic analysis (\hyperref[para:allocsize]{Allocation Sizes}) comes to our aid!

We bake the dynamic analysis information (e.g., $\aliasinfo, \sizeinfo$) into an Argument Construction API.
These API functions convert C objects into \textit{auxilliary} Rust objects that adhere to size/aliasing relations between arguments.
API functions are exposed as tools to the LLM. The LLM can then write a wrapper \translateargs{} function that coerces the auxiliary Rust objects into the final objects that match the function signature.

\begin{table}
\centering
\begin{tabular}{c|c|c}
\hline
\textbf{Function} & \textbf{(input) \clanguage Type} & \textbf{(output) Auxilliary \rustlang Type} \\ \hline
\texttt{translate\_prim\_single} & $\cct{int}, \cct{char}, \cdots$ & $\rustct{u32}, \rustct{i8}, \cdots$ \\
\texttt{translate\_prim\_vec} & $\cct{int*}, \cct{char*}, \cdots$ & $\rustct{Vec<u32>}, \rustct{Vec<i8>}, \cdots$ \\
\texttt{translate\_string} & \cct{char*} & \rustct{String} \\
\texttt{translate\_struct} & \cct{struct T*} & \rustct{Rc<RefCell<Struct>>} \\ \hline
\texttt{translate\_1d} & \cct{T*} & \rustct{Vec<Rc<RefCell<RT>>>} \\
\texttt{translate\_2d} & \cct{T**} & \rustct{Vec<Vec<Rc<RefCell<RT>>>} \\ \hline
\end{tabular}
\caption{Argument Construction API Functions: functions to translate primitive singletons and arrays, strings, and structs. For n-D arrays, we provide functions to translate them into (possibly multi-dimensional) \rustc{Vec} objects. In these cases, we preserve aliasing matching C using \rustc{Rc<RefCell<RT>>} (Rust smart pointers) - here the \rustct{RT} is the Rust type that the C type \cct{T} maps to.}
\label{tab:aca}
\end{table}

We provide the API functions in Table \ref{tab:aca}.
These include functions translating singleton primitive objects (e.g., integers), strings, and structs.
For pointers referencing n-D arrays, we provide functions to translate them into (possibly multi-dimensional) \rustc{Vec} objects.
For the \cc{sum} and \cc{caps} functions in Fig. \ref{fig:signature}, the \llm{} can use the \texttt{translate\_1d} and \texttt{translate\_string} API functions in \translateargs{}.
The API functions use the dynamic analysis under the hood to infer the size of the \rustc{Vec} and the length of the string.
The API functions also use the dynamic analysis information (e.g., $\aliasinfo, \sizeinfo$) to translate the aliasing properties between C objects to the Rust objects.
For example, if two C struct pointers \cct{struct s * a, b} reference the same struct object, 
the translation API will create a single Rust struct (\rustc{rust_s}) object and return two \rustc{Rc<RefCell<rust_s>>} references to it. Below, is an example \llm{}-generated \translateargs{} for \cc{sum}. Notably, the APIs are helpful high-level primitives, while \llm{} is allowed to write arbitrary Rust when constructing Rust objects (e.g., converting \rustc{Vec<i32>} to \rustc{Vec<u32>} in the code below).

\begin{lstlisting}[style=RStyle, basicstyle=\fontencoding{T1}\fontfamily{lmtt}\scriptsize]
pub unsafe fn translateArgs_sum(a: *mut c_int) -> Vec<u32> {
    let rust_a = translate_1d::<c_int>(
        a as *mut c_void
    ).expect("Failed to translate a")
    .into_iter()
    .map(|x| x as u32).collect::<Vec<u32>>(); // Convert from Vec<i32> to Vec<u32>

    rust_a
}
\end{lstlisting}

\subsubsection{Equivalence Testing}
\label{subsubsec:eqtestgen}

With a valid \translateargs{}, we can execute $f_C$ and $f_R$ on equivalent inputs.
Here, we observe that functional equivalence between $f_C$ and $f_R$ can be elegantly demonstrated through a commutative relationship centered on the \translateargs{} function.
%
\begin{equation*}
\begin{tikzcd}
    \ppre_\clanguage \arrow{r}{f_C}  \arrow[swap]{d}{\translateargs} & (\ppre_\clanguage, \ppost_\clanguage) \arrow[]{d}{\translateargs} \\%
    \ppre_\rustlang \arrow{r}{f_R}& (\ppre_\rustlang, \ppost_\rustlang)
\end{tikzcd}
\end{equation*}
%
That is, for $f_C$ and $f_R$ to be equivalent, directly executing $f_C$ and then translating its output must produce the same result as first translating the inputs and then executing $f_R$. 
Consequently, we ask an \llm{} to generate an equivalence test that performs this validation logic and \rustc{assert} that each of the C and corresponding Rust outputs agree. We limit the assertions to perform \textit{value-based equivalence} only. 
Below is a simple example of an equivalence test for \cc{caps} in Fig. \ref{fig:signature}:\\

\begin{lstlisting}[style=RStyle, basicstyle=\fontencoding{T1}\fontfamily{lmtt}\scriptsize]
pub unsafe fn eqtest_caps(filename: &str) -> bool {
    let c_pre_arg0 = load_pre_args_from_json(filename);             // Load C inputs
    let mut rust_string = translateArgs_caps(c_pre_arg0);           // Translate to Rust input

    let c_post_arg0 = load_post_args_from_json(filename);           // Load C outputs
    let expected_rust_string = translate_args_caps(c_post_arg0);    // Translate to expected Rust output
    
    caps(&mut rust_string); // Call Rust function with translated input

    assert_eq!(rust_string, expected_rust_string, "Rust string does not match expected Rust string");

    true
}
\end{lstlisting}

\subsection{Multi-round \& Rejection Sampling}
\label{subsec:irs}

Overall, with this pipeline, 
specification mining
(\S\ref{subsec:dynamicspec}) $\rightarrow$ 
code generation (\S\ref{subsec:codegen}) $\rightarrow$
testing (\S\ref{subsec:testgen}), we can identify whether a generated $f_R$ is a valid translation of $f_C$.
We use multiple rounds of this pipeline with error feedback to correctly translate all functions.

\paragraph{Rejection Sampling} As mentioned in \secref{subsec:appoverview}, we make use of sampling to scale \llm{} inference and verify generated solutions. We use this rejection sampling approach at each pipeline module that invokes an \llm{}, as illustrated in Figure \ref{fig:rejsampleloop}. For each module, we first sample a set of solutions, i.e., rust translations for \codegenerator{}, \translateargs{} for \argtranslator{}, and equivalence tests for \eqtester{}. We then use either compilation or execution as the signal to filter out bad generations and pass the filtered set to the next module.

\paragraph{Error Feedback} Finally, failing equivalence tests are used as error feedback. 
However, since these functions can produce large outputs, we perform a \texttt{diff} operation between the true and desired output to isolate the discrepancy.
We provide these diffs and assertion messages to the \llm{} and ask it to regenerate the (incorrect) function.
Consider an anecdotal example for equivalence-based feedback from our \zopfli{} case study (\secref{subsec:zopfli}). Below, the \cc{BoundaryPMFinal} function grabs the \cc{next} field from a \cc{pool} of nodes (left). However, in the first round, the \llm{} interprets \cc{node->next} as incrementing the index field (right), corresponding to the \cc{next} iterator in the C code.
\noindent
\begin{center}
\begin{minipage}{0.40\textwidth}
\begin{lstlisting}[style=CStyle, xleftmargin=0cm, basicstyle=\fontencoding{T1}\fontfamily{lmtt}\tiny]
if (lcount < nsymb && sum > leaves[lcount].weight) {
  Node *newch = pool->next;
  // ^^^^^^^^^^^^^^^^^^^ this line
  Node *oldch = lists[index][1]->tail;
  ...
}
\end{lstlisting}
\end{minipage}
\hspace{0.75cm}
\begin{minipage}{0.45\textwidth}
\begin{lstlisting}[style=RStyle, xleftmargin=0cm, basicstyle=\fontencoding{T1}\fontfamily{lmtt}\tiny]
let newch = pool
  .next.get(pool.index)
  .expect("Pool has no more nodes")
  .clone();
// Move to next free node in the pool.
pool.index += 1; 
// ^^^^^^^^^^^^ this line
\end{lstlisting}
\end{minipage}    
\end{center}
Here, our testing approach allows deserializing iterators with index information and catching bugs in the \llm{}'s code, which unnecessarily updates the iterator's position. As a result, providing error feedback allowed the \llm{} to correct the code using our diff-based localization.

\begin{figure}[t]
\centering
\begin{tikzpicture}
  \node (A) at (0, 0) {$f_C$};
  \node (B) at (2.5, 0) {$\left\{ f_{R_i} \right\}_{i=1}^{N}$};
  \node (C) at (6, 0) {$\left\{ f_{R_i} \right\}_{i=1}^{N' \leq N}$};
  \node (D) at (11, 0) {$\left\{ 
\begin{matrix} 
(f_{R_i}, t_{ij})
\end{matrix}
\right\}_{ij=1}^{M = N' \times K}$};
  \node (E) at (11, -2) {$\left\{ 
\begin{matrix} 
(f_{R_i}, t_{ij})
\end{matrix}
\right\}_{ij=1}^{M' \leq M}$};
  \node (F) at (6.5, -2) {$\left\{ 
    \begin{matrix} 
    (f_{R_i}, t_{ij}, eq_{ijk})
    \end{matrix}
    \right\}_{ijk=1}^{S = M' \times K}$};
  \node (G) at (1, -2) {$\left\{ 
(f_{R_i}, t_{ij}, eq_{ijk}) \right\}_{ijk=1}^{S' \leq S}$};

  \draw[->] (A) -- node[above] {\text{CodeGen}} (B);
  \draw[->, dashed] (B) -- node[above] {\text{Compile}} (C);
  \draw[->] (C) -- node[above] {\text{TranslateArgs}} (D);
  \draw[->, dashed] (D) -- node[left] {\text{Execute}} (E);
  \draw[->] (E) -- node[above] {\text{EqTest}} (F);
  \draw[->, dashed] (F) -- node[above] {\text{Execute}} (G);
\end{tikzpicture}
\caption{Illustration of the rejection sampling approach used to translate a C function $f_C$.
At each stage, the pipeline oscillates between a \textit{generation} ($\rightarrow$) and \textit{verification} ($\dashrightarrow$) step that filters valid generations. 
The process starts by the \llm{} generating $N$ candidate Rust functions, from which compilation filters valid programs ($N' \leq N$). Then the \argtranslator{} samples $K$ \rustc{translate_args} functions for each compilable Rust function, resulting in $M$ function-translator pairs. Execution filters these to $M'$ valid pairs, which the \eqtester{} expands into $S$ triples by sampling $K$ equivalence tests per function-translator pair. A final execution phase filters these to $S'$ triples, yielding successfully translated and equivalent Rust functions.
}
\label{fig:rejsampleloop}
\end{figure}

\section{Implementation}
\label{sec:implementation}
Following, we describe the implementation details of the various components of our approach.

\subsection{\slicer{}} 
We implement our program slicer using \clangseventeen{} that consumes an entire C codebase, analyzes the definitions and usages, and builds a topologically sorted dependency graph of all the code elements in the C program. Additionally, we implement an iterator over the graph that progressively traverses and yields code elements in dependency order.
By performing a topological sort on this graph, we can identify an incremental translation unit and determine its dependency slice (the set of units it depends on).
Additionally, we incorporate extra dependency edges between function types and the functions matching those types, as inferred via dynamic analysis.

\subsection{\specminer{}} 
\specminer{} module of our pipeline is responsible for collecting I/O specifications for \clanguage{} functions.
These I/O are later used for inferring various property specifications such as nullability related to the function.
Since \clanguage{} does not support reflection and even primitive information such as pointer sizes and types are not directly available, we use dynamic analysis to collect this information.
The \specminer{} module of our pipeline is responsible for collecting input/output (I/O) specifications for \clanguage{} functions.
These I/O specifications are later used to infer various function properties, such as nullability and aliasing.
Since \clanguage{} does not support reflection, and even primitive information like pointer sizes and types are not directly available, we employ dynamic analysis to collect this information. 
In particular, we implement an allocation tracking \llvm{} pass (using \llvmfourteen{}\footnote{Note that opaque pointers in more recent \llvm{} versions do not allow easily accessing type information.}) to collect the bounds and type information for all allocations in the program. 
We then implement a \cpplang{} runtime to access this information for serializing the \clanguage{} function arguments and outputs. 
Because C functions can be stateful and may modify their arguments, we serialize the arguments both before and after function execution along with the return value.

\subsection{\codegenerator{}} 
We implement our \rustlang{} code translation module using an \llm{} based \textit{parallel} sampling and filtering approach.
In particular, for a given \clanguage{} translation unit, we use the \slicer{} to collect its immediate parents. 
We then prompt the \llm{} with the \clanguage{} source code elements, the corresponding \rustlang{} elements already translated, and the dynamic analysis annotations (such as nullability and aliasing information about arguments) to generate the translated \rustlang{} code.

We use parallel sampling and a multi-turn repair approach to generate the \rustlang{} translations.
In particular, we sample multiple candidate translations and use \rustlang{} compiler with \rustc{#![forbid(unsafe_code)]} directive to identify safe compiling translations. 
We keep the compiling translations and prompt the \llm{} with the compile-error feedback to refine the failing translations (with backoff to explore more candidate translations).
We repeat this procedure across multiple rounds to generate a capped number of minimum compiling translations.
This approach allows us to generate a diverse set of translations efficiently which can allow finding a \textit{correct} (that is passing the \eqtester{}) translation faster.
Particularly, we can defer needing execution feedback based multi-turn sampling after \eqtester{} if any one of the candidate translations passes the \eqtester{}.

\subsection{\argtranslator{}} 
We use \rustc{bindgen} to generate \rustlang{} function signatures from the \clanguage{} function signatures and employ \rustc{FFI} to access the C functions and their arguments.
We then use \llms{} to programmatically map the types and arguments between the \clanguage{} and \rustlang{} functions. 
Specifically, for a given \clanguage{} and its candidate  \rustlang{} translation, translation, we provide the \llm{} with the \textit{aligned} \clanguage{} and \rustlang{} function contexts and ask it to generate a program that maps the \clanguage{} arguments to appropriate \rustlang{} arguments.
We use successful execution of the generated translation program (that is, we can load and map the \clanguage{} arguments to \rustlang{} arguments) as a filter to select the candidate \rustlang{} functions.

\subsection{\eqtester{}}
For a given \clanguage{} function, its candidate translated \rustlang{} function, and the corresponding type alignment function, we prompt the model to generate a test function which checks whether the \clanguage{} and \rustlang{} functions are equivalent.
We perform value-based equivalence checks by implementing the \rustc{PartialEq} traits to \rustlang{} types, enabling the use of the built-in equality operator to assert equivalence (performing nested equality comparisons internally).
We use these generated tests to filter the candidate \rustlang{} functions that successfully compile and pass the equivalence tests. 
If none of the candidate functions pass the tests, we provide the model with execution error feedback and iteratively repair the translation over multiple rounds.

\noindent We run our experiments on a $96$ core Intel Xeon machine with $720$ GB of RAM.
We use \textsc{o1-preview} and \textsc{o1-mini} to perform all the steps unless specified otherwise.

\section{Experiments}
\label{sec:experiments}

In this section, we discuss two experimental results.
First, in \secref{subsec:urlparser}, we present a case study on translating \urlparser{}, a modest \clanguage{} program comprising over $400$ lines, which was studied in prior work~\cite{li2024translating}. 
Next, in \secref{subsec:zopfli}, we present our results from translating \zopfli{}, an optimized \clanguage{} compression program with over $3000$ lines.

\subsection{\urlparser{} Case Study: Motivating Dual Translations}
\label{subsec:urlparser}

We translate the \urlparser{} program~\cite{urlparser}, a \clanguage{} application that parses URLs and extracts their components (protocol, domain, path, query, and fragment) from a URL string.
To gather deeper insights, we run our approach with human intervention to identify failure models and solutions.
This program has been used in prior work~\cite{li2024translating}, where the authors applied existing \llm{}-based approaches, namely Flourine~\cite{eniser2024towards} and VERT~\cite{yang2024vert}, and reported challenges in generating correct translations.
In particular, they faced difficulties in program decomposition, where inconsistent translations of decomposed functions resulted in compilation errors.

\SubParagraph{Incremental sampling and filtering alleviate challenges}
We use our \slicer{} and \codegenerator{} modules to incrementally generate compilable translations.
%
By providing dependency context to the \llms{} during translation and leveraging the multi-turn compiler feedback-based \codegenerator{} module, we can generate a compilable translation for the \urlparser{} program.

\SubParagraph{Compilable code translations do not suffice}
However, at this stage, we do not utilize execution filtering when generating individual function translations. 
We run tests for the top-level \rustlang{} \rustc{url_parse} method and find that none of the candidate translations passes test test case.

\SubParagraph{Testing alleviates challenges}
To address this issue, we semi-manually construct tests for individual functions in the \urlparser{} program and use these tests as additional filtering criteria during translation. 
These tests are translated from the \urlparser{} repository by prompting LLMs with \clanguage{} tests and incorporating manual interventions to ensure test correctness. 
By using test execution as additional feedback, we can generate a correct translation for the top-level \cc{url_parse} function, thereby demonstrating the effectiveness of our dual translation approach.

\subsection{\zopfli{} Translation}
\label{subsec:zopfli}

hWe translate the C program for the \zopfli{} compression algorithm~\cite{zopfli}, which is known to achieve superior compression ratios by extensively optimizing the compression process. 
The codebase consists of over three thousand lines of C code (excluding comments), comprising ninety-eight functions and ten structs, and spans over twenty-one files. 
It provides a challenging testbed for evaluating our approach due to the diversity of the source code constructs, including heap-based data structures (such as linked lists and array iterators), function pointers, and \cc{void*} arguments while still agreeing with our assumptions about acyclic data structures, concurrency, and lack of \cc{void*} abuse (type punning).
The C code implements nuanced algorithms, including LZ77 compression, Huffman encoding, and block splitting.
Following, we detail our implementation and results.

\subsubsection{Implementation}
We use the \cc{unifdef} tool~\cite{unifdef} to preprocess the codebase and standardize the \cc{\#ifndef} preprocessor configuration options. 
This process addresses a combination of system configurations (operating system, compiler) and optimization settings implemented in \zopfli{}. 
Next, we run \sysname{} on \zopfli{}, using \cc{ZopfliDeflate} as the top-level entry point interfacing with the core deflate algorithm. 
This function receives the input string to be compressed, along with \cc{ZopfliOptions} configurations that determine how block splitting is performed. 
We semi-automatically collect \numzopflitests{} test inputs for the \cc{ZopfliDeflate} function. 
These inputs achieve \clinecoverage{}\% line coverage and \cbranchcoverage{}\% branch coverage on the original C program.

We use \specminer{} to collect I/O and property specifications for every function in the \clanguage{} code using the above-mentioned tests dynamic allocation analysis. 
Next, we iterate on the translation units in dependency order and run our pipeline to generate \rustlang{} translations for each unit and test for equivalence.
Since we do not have equivalence tests for non-function translation units, such as structs, global assignments, and macros, we manually verify translation correctness.
In particular, we manually construct the appropriate \rustlang{} structs and check other global assignments or macros as necessary.
This process discovered a bug in the \cc{ZOPFLI_APPEND_DATA} macro where \llm{} translation did not handle a corner case in the original code which we repaired manually.

\subsubsection{Results}
We run our approach and translate \zopfli{} in about $15$ hours costing about $2500\$$\footnote{We believe the translation can be done within $1500\$$ with better hyper-parameters (choice of models, number of samples) and more robust infrastructure of our pipeline. Given the high cost, we do not study pipeline optimizations at this time.}.

\Paragraph{Equivalence Test Suite}
To ensure that our translation is correct, we collect a comprehensive test suite for the \cc{ZopfliDeflate} function comprising $\numzopflitestseval{}$ compression inputs ranging between $1e1$ and $1e7$ characters (having a combined size over $20$GB).
We measure coverage of our larger tests and find that they achieve $\cevallinecoverage\%$ line coverage and $\cevalbranchcoverage\%$ branch coverage.
Note that we identified a considerable portion of uncovered code regions (example $5.9$\% of the branches) as error regions (like \cc{assert} statements, \cc{exit} branches) and thus supposed to be unreachable. 
We use these to check equivalence in compression ratios between the \clanguage{} and \rustlang{} programs. 

\Paragraph{Test Results and Repair}
We observed run-time exceptions for a fraction of our tests.
We use the failing equivalence test and our \specminer{} module to identify the \textit{innermost} function that fails the per-function I/O specification and \textit{restart} our pipeline from the failure point.
Particularly, we identify the failure in \rustc{zopfli_block_split_lz77} function and prompt the model to repair the failing test. 
This bug was caused by less comprehensive tests used in the previous run of the pipeline, and we find that \llm{} could easily fix this with a simple and local change: slicing the \rustc{splitpoints} \rustc{Vec} (shown in Figure~\ref{fig:custom-repair}). 
Note that this repair does not follow the greedy translation principle from our approach (since functions dependent on \rustc{zopfli_block_split_lz77} already exist) and can have cascading effects.
However, we found that the bug mentioned above and the corresponding fix were local. After performing this repair, we successfully passed our entire test suite, thus providing high confidence in the correctness of our translated code.
\begin{figure}[h!]
    \centering
\hspace{-0.5cm}
\begin{minipage}{0.44\textwidth}
\begin{lstlisting}[style=RStyle, xleftmargin=0cm, basicstyle=\fontencoding{T1}\fontfamily{lmtt}\scriptsize]
    if !find_largest_splittable_block(
        lz77.size,
        &done,
        splitpoints,
        &mut new_lstart,
        &mut new_lend,
    ) {
        break; // No further split will probably reduce compression.
    }
\end{lstlisting}
\end{minipage}
\hspace{0.8cm}
\begin{minipage}{0.45\textwidth}
\begin{lstlisting}[style=RStyle, xleftmargin=0cm, basicstyle=\fontencoding{T1}\fontfamily{lmtt}\scriptsize]
    let found = find_largest_splittable_block(
        lz77.size, &done,
        &splitpoints[..*npoints],
        &mut lstart, &mut lend,
    );

    if !found {
        break; // No further splits will likely reduce compression.
    }

\end{lstlisting}
\end{minipage}
\caption{
We discovered some runtime exceptions when running our equivalence test suite.
We used our \specminer{} approach to find the \textit{inner-most} function which failed the \textit{extracted} function-level I/O specification and re-run the pipeline from that point.
We found that \llm{} was able to successfully repair the rust code by appropriately performing the slicing on \rustc{splitpoints} \rustc{Vec} (line 4 on the left and line 3 on the right).
Moreover, after applying this local change, our \rustlang{} code passes the entire test suite of $\numzopflitestseval{}$ tests.
}
\label{fig:custom-repair}
\end{figure}

\Paragraph{Pass-rates}
We study the quality of the translation candidates across different functions. 
Particularly, we measure two metrics -- compilation pass rate and execution pass rate. 
Compilation pass rate is computed as the number of translations that compile against the total number of translations generated.
The execution pass rate is computed as the number of translations that pass the equivalence test compared to the total number of compiling translations.

\begin{figure}[!t]
    \centering
    \includegraphics[width=0.65\linewidth]{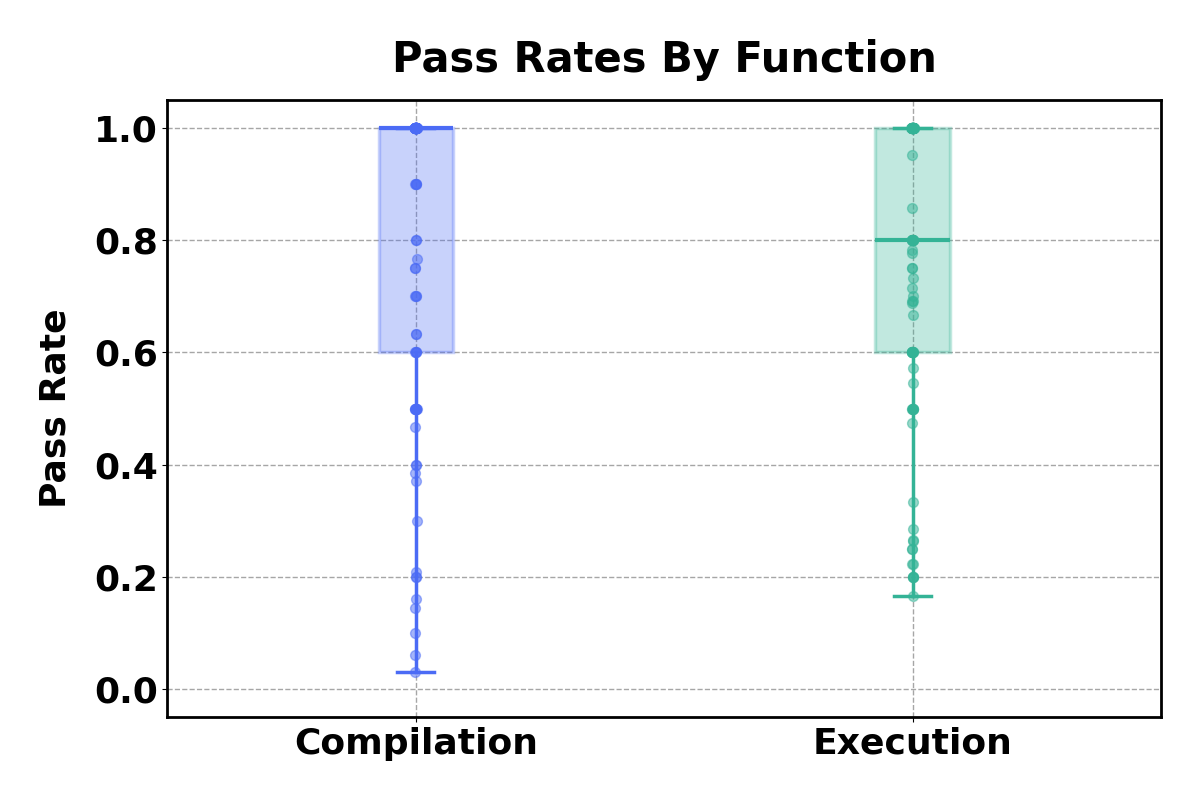}
    \vspace{-5pt}
    \caption{\textbf{Box plot demonstrating pass rates for Code 
 Compilation and Test Execution.} We measure the per-function pass rates for compilation and test execution. Note that for test execution, we use the compiling solutions as the set of submitted solutions (denominator when computing pass rates).}
    \label{fig:passrates}
    \vspace{-8pt}
\end{figure}

Figure~\ref{fig:passrates} depicts a box plot highlighting the mean, median, and distribution of pass rates over different function samples. 
First, we observe that the mean pass rates are high, ranging over \textbf{75}\% for both getting a compilable solution and an executable solution. 
This indicates that our approach can generate high-quality translations for a large fraction of the functions somewhat easily (with high pass rates).
However, this distribution is quite skewed with a long tail of \textit{challenging} function with low pass rates (below 20\%).

\subsubsection{Analysis} Next, we analyze the quality and efficiency of translations. 

\Paragraph{Qualitative Analysis}
Our translated Zopfli program ranges over 7k lines (including overly verbose comments and docstrings). 
Interestingly, we observe that \llms{} eagerly modify the C program's logic (usually in the spirit of optimizing or simplifying), often leading to bugs.
We address these issues by prompting the models with strict warnings to limit such behaviors.
For example, since \clanguage{} does not provide native support for data structures like vectors, it maintains additional attributes like the length of an array to mimic the functionality. 
\llms{} are eager to write \rustlang{} code that mitigates using external length (and using \rustc{Vec} \rustc{length}) but do so inconsistently causing bugs.

We also note \llms{} struggle to generate correct (or even compiling) code for long functions, often missing or skipping out the translation of some intermediate component. 
One particularly nefarious case for this originates from the difference in \clanguage{} and \rustlang{} \cc{for} loops. 
Specifically, in \clanguage{}, you can modify the loop iterator variable inside the loop body. \rustlang{} does not provide this option and the translation needs to map the \cc{for} loop to \cc{while} loop, handling incrementing the iterator at all loop \cc{continue} statements, often not managed correctly in long programs by \llms{}.

\Paragraph{Efficiency}
We next compare the efficiency of the original and translated programs on two test suites -- first with six strings formed by repeating \cc{'a'} varied number of times (between $1e1$ and $1e6$ times) and second with twelve random strings of varying lengths (between $1e3$ and $1e6$ characters).
Additionally, we compare the execution times under default compilation configuration (`\cc{gcc -O0 -g}' and `\cc{cargo debug}') and optimized compilation configuration (`\cc{gcc -O3}' and `\cc{cargo build --release}') for both the C and Rust implementations.

We find that the translated Rust code is slower, with larger slowdowns, in the default compilation configuration compared to the optimized compilation configuration. 
Next, even with the optimized compilation configuration, the Rust code is up to $3.67$x slower than the C code.
We study this further by comparing the flamegraphs for the default compiled versions. 
We estimate a considerable fraction of slowdown from time spent on \rustc{Vec} allocations and bound checks in \rustlang{}. Additionally, we identified  \cc{ZopfliUpdateHash}, a function that modifies data structure containing large arrays, as a large contributor to this slow-down.
Note that \rustlang{} might already be optimizing some of these issues, and understanding performance differences between the optimized versions would require more careful time measurements.
\begin{table}[ht]
    \centering
    \caption{Performance Comparison of C and Rust Implementations}
    \begin{tabular}{l|c|c|c|c|c|c}
    \hline
    \multirow{2}{*}{Input Type} & \multicolumn{3}{c}{Default} & \multicolumn{3}{c}{Optimized} \\
    \cline{2-7}
     & C & Rust & Slowdown & C & Rust & Slowdown \\
    \hline
    Repeated Input & 0.17 & 1.56  & 8.9x & 0.040 & 0.059 & 1.47x \\
    Random Input & 0.242 & 3.336 & 13.8x & 0.094 & 0.330 & 3.67x \\
    \hline
    \end{tabular}
    \label{tab:performance_comparison}
\end{table}

\subsubsection{Ablations}
Following, we describe ablations for our approach.

\Paragraph{Choice of Models}
Recall (from ~\secref{sec:implementation}) that we use \textsc{O1} models for our experiments.
We also experiment with \textsc{GPT-4O} models to perform translation. 
Since \textsc{GPT-4O} models are less powerful, we increase the sampling budget to collect more candidate translations. 
Additionally, since we discovered that \textsc{GPT-4O} is insufficient for performing execution-feedback-based repair operations, we fell back to \textsc{O1} models for repair.
We run our pipeline, passing all intermediate tests, and generate an alternative translation for \zopfli{}.
Notably, using significantly cheaper \textsc{GPT-4O} models, we can generate a correct translation for \zopfli{} in $\$<800$.
However, the generated translation is noticeably slower than the original \clanguage{} implementation and crashes on some long test inputs.
We conclude that \textsc{O1} models produce more reliable and higher-quality translations and hope to improve the efficiency of our pipeline by using robust combinations of \textsc{O1} and \textsc{GPT-4O} models in future work.

\Paragraph{Choice of using dynamic analysis and Testing}
Recall that we use dynamic analysis to construct intermediate tests and incrementally generate correct code.
As an ablation, we remove the testing module directly to generate code without any intermediate filtering except compiler checks, resembling the approach from ~\cite{shiraishi2024contextaware}.
Using this approach, we can successfully generate a compiling version of \zopfli{} using \llm{} queries within $100\$$.
However, we find that the generated code does not run even for trivial inputs such as \rustc{"Hello, World"} and crashes with \rustc{index out of bounds} error. 
This aligns with our findings from Figure~\ref{fig:passrates} where compiling code solutions achieves only $80\%$ median pass-rate with long-tail achieving $<20\%$ pass-rate.


\section{Related Work}
\label{sec:related}

\subsection{C-to-Rust Interest and Applications}
C to Rust full code translation has recently garnered much interest owing to memory safety vulnerabilities \cite{eternalwar} in C, strong typing and ownership in Rust, and the relative ease of combining C and Rust through Clang compilation toolchains, and the C-Rust FFI (Foreign Function Interface).
In particular, the NSA (US National Security Agency) has strongly advocated migration to Rust, and DARPA has launched the TRACTOR program \cite{darpa} aimed at automating translation.

We based our main case study on \zopfli{} \cite{zopfli}, a high-performance compression library by Google.
We chose \zopfli{} because (a) compression libraries form a project milestone of the DARPA TRACTOR program, and (b) \zopfli{} has been manually translated to Rust \cite{rust-out-your-c} with the experience well-documented.
We note that our auto-translation radically differs from the handwritten implementation (e.g., in terms of code architecture, organization, and Rust features used).
There is also emerging user-study research comparing human expert-translated and auto-generated translations (e.g., using LLM-based techniques) \cite{li2024translating}.

Translation from C to Rust is considered especially relevant for sensitive codebases such as cryptographic and network libraries and OS kernel modules/drivers.
However, the opinion is still divided on \textit{how much} and \textit{what part} of the source C codebase to migrate.
For example, \citet{rustforlinux} makes the case that certain low-level utilities (e.g., drivers) are best left in C since migration can lead to more subtle bugs (e.g., issues with memory mapping/address spaces).

\subsection{Automated C-to-Rust Translation}
\label{subsec:ctorust}

%
We provide a summary of C-to-Rust translation approaches in Table \ref{tab:comparison}.
The work on C-to-Rust translation closest to ours is \cite{shiraishi2024contextaware}. They, too, perform \llm{}-driven C-to-Rust translation. 
However, unlike us, they focus on sampling alone, without intermediate testing (for internal functions).
As a result, they achieve lower accuracy on the validation tests.
One of our key insights is that incremental translation combined with internal function tests can result in major accuracy gains for moderate-large (multi-file/function) codebases.
Accordingly, we enable internal equivalence tests by collecting C I/O examples (\secref{subsubsec:c-io-spec}) and translating them to Rust (\secref{subsubsec:eqtestgen}).

VERT \cite{yang2024vert} also uses LLMs to generate Rust code from multiple source languages (including C) using an MSWasm-based (memory-safe WebAssembly) testing infrastructure. 
However, they also do not perform incremental translation/internal function tests.
Further, they evaluate their approach on much smaller (<200 LoC) code sizes.
Related to this is \cite{eniser2024towards} which performs \llm{}-driven translation with differential testing and repair (when top-level tests fail). They evaluate the performance of different LLMs with respect to accuracy and repair feedback strategies.
However, they do not perform intermediate tests and only achieve limited success on small benchmarks.

Other C-to-Rust translation approaches include c2rust \cite{c2rustgalois}, which translates C to \textit{unsafe} Rust making heavy use of C-like constructs and C-types through C-Rust FFI.
Beyond \rustc{unsafe}-ness, this results in artificial and less interpretable code.
Immunant and Galois aim to extend c2rust to produce \textit{safer} Rust \cite{immunant}.
They, too, identify a use-based (semantic) type correspondence between C pointers and Rust reference types, similar to our Semantic Type Alignment (ref. \secref{subsec:semtypealign}).
Further, very recently, they too have explored nullability analysis (as their blog \cite{galois} indicates), albeit using \llms{}. We, on the other hand, use classical dynamic analysis (\secref{subsec:dynamicspec}), which we believe is more predictable/reliable.
CROWN \cite{zhang2023ownership} takes the output of c2rust and performs best-ownership effort analysis and rule-based rewriting to convert it to \rustc{safe} Rust.
While their approach scales well, usage of \rustc{unsafe} remains.

\subsection{\llm{}-driven Code Translation}
\llm{} driven code translation has received considerable interest in recent years. 
These works have explored  training \llms{} for code translation ~\cite{RoziereLachaux2020, RoziereZhang2022, structcoder}, leveraging compiler representations~\cite{szafraniec2022code}, and directing prompting based approaches~\cite{tang-etal-2023-explain, jiao2023evaluation,jana2023attention, yin2024rectifiercodetranslationcorrector}.
However, these works have primarily focused on translating simple competition or interview-style programming problems. 
\citet{cassano2024knowledge} used a rule-based test translation to translate small (single-function) program snippets from high-resource to low-resource languages.

%
%
Two recent works use \llms{} to perform repository-level translation with test-based validation.
Particularly, \citet{ibrahimzada2024repositorylevel}
study Java to Python and \citet{zhang2024scalable} study Go to Rust translation.
While our work shares themes of incremental translation and test-based validation with these works, key aspects of our approach are unique to C-to-Rust translation as we now discuss.

\textbf{Semantic execution information aids code/test generation.}
We observe that C constructs such as dynamic allocations, pointer-based direct memory accesses, aliasing, and nullability make interpreting (and hence translating) C code challenging.
Analyses that extract these properties, and exposing them to the \llm{} can enable more accurate, and hence, scalable translation.

\textbf{Dynamic analysis enables collecting semantic information.}
%
While semantic execution information and function I/O aids code generation and testing respectively, extracting these is challenging due to weak typing and pointer casts in C.
We tackle this challenge this by developing a dynamic specification mining infrastructure (\secref{subsec:dynamicspec}) that collects this information.

While Java-to-Python and Go-to-Rust have their own challenges, memory management patterns translate fairly directly. 
Furthermore, these languages have rich standard libraries with parallel high-level constructs, unlike \clanguage{}'s minimal set.
Finally, our \sysname{} approach to translation attempts to scale \llm{} inference using techniques such as repeated sampling and execution-feedback-based repair to achieve completely valid and equivalent translations.


\subsection{Agentic Code Generation}
There is a tremendous amount of research on \llm{}-driven code generation. 
Here, we highlight relevant works on agentic code generation and direct readers to appropriate surveys on the topic~\cite{jiang2024survey,liu2024large}.
\llm{} based programming agents have been gaining popularity with benchmarks~\cite{r2e,swebench,lcb,shi2024language} and agentic approaches for solving them~\cite{sweagent,agentless,openhands,wang2024planningnaturallanguageimproves}.
Our approach indeed forms a non-agentic pipeline for solving challenging long-horizon code (translation) problems.
\citet{bairi2024codeplan} proposed CodePlan which performs repository-level code editing using program analysis (dependency graphs).
Recent works have also focused on inference-optimal compute usage~\cite{wu2024inference,snell2024scaling,zhang2024scaling,damani2024learning} to optimize inference costs, which is orthogonal to our work but nonetheless an important aspect. 
Finally, we believe many ideas from traditional code generation work will transfer to the code translation domain~\cite{chen2022codet,ni2023lever,zelikman2022parsel,jain2023llm, olausson2023self,chen2023teaching,zhang2023coder,inala2022fault,Jigsaw,shinn2024reflexion}.

\section{Discussion}

\subsection{Challenges and Future Work}
\label{subsec:challenges}
Now we enumerate some unalleviated challenges we plan to address in future work.

\Paragraph{Golden Translation Specification}
Note that we require that the generated Rust program satisfy (a) equivalence on test inputs and (b) not use \rustc{unsafe} Rust.
In general, requiring no-\rustc{unsafe} may be an overly strong constraint.
Some unsafe executions may be acceptable if they are unexploitable, i.e., user-facing functions cannot be invoked in a way that leads to an attack. Low-level C code often uses (safely abuses) such behaviors. Prohibiting unsafe constructs in such cases can lead to artificial Rust code that may suffer from other bugs (e.g., due to subtle differences in C/Rust kernel allocation methods \cite{rustforlinux}). Thus, a golden (ideal) specification for C to
Rust translation would strike a balance by only forbidding exploitable unsafe behaviors.

Determining which usages of unsafe behavior are unexploitable is extremely challenging, and requires a combination of security (attacker) modelling and program analysis. 
A more lightweight approach would be to allow the user to annotate certain C functions, e.g., as \texttt{NO\_TRANSLATE}, meaning that the said function should be kept in C and integrated with the rest of the translated Rust codebase through the FFI (Foreign Function Interface).
We believe that our infrastructure can support this with relatively low effort.
In general, future work may take a more nuanced approach to thread the needle between ensuring C-equivalence and enforcing \rustc{safe} Rust.


\Paragraph{Translation performance, Cost and Speed}
Our approach relies on \llm{} capabilities to perform the translation and uses program analysis and testing to scaffold the process, improving reliability. 
Here, we identify that while \llms{} can be unreliable and make mistakes, we can make them reliable by sampling many candidate translations and filtering them using tests. 
Additionally, we decompose the translation process into smaller parts, which further reduces the complexity of the translation, improving the reliability of \llm{} based translation.
Thus, our performance is bounded by \textsc{Pass@Any} of \llms{}--often considerably higher than \textsc{Pass@1}~\cite{llmmonkeys}.

This increases translation time and cost, which becomes crucial as we scale to larger programs.
Future work should explore how to orchestrate the translation \textit{search} process to reduce the cost of translation following \textit{agentic} approaches in existing \llm{} for code literature~\cite{wang2024planningnaturallanguageimproves}.

\Paragraph{Dependency Ordered Translation}
\phantomsection{}
\label{disc:deporder}
We translate the code in a greedy topo-sorted order of the program dependency graph (\secref{sec:approach}).
The greedy algorithm reduces to combinatorial search complexity but may lead to sub-optimal translations.
Particularly, as described in \secref{sec:approach}, accurately translating structs requires global context of how the struct is used in the program (providing information about how to refactor the struct fields).
We currently resolve this by manually translating the structs and aim to explore how \llms{} and usage analysis can guide the translation of structs in the future.
Next, while translating functions, we identify various dynamic analysis-based property annotations (nullability, aliasing, etc.) that can guide the translation process without requiring the entire program context.
However, this may again lead to sub-optimal translations due to \textit{bad} translation choices made early in the translation process. 
For example, translating a pointer argument to an array early on might lead to conflicts when later translations require the argument to be a vector. This would add unnecessary type casts, reducing program efficiency.
Future work should explore how to solve this global consistency problem in the translation process.


\Paragraph{Test-based Equivalence: Incomplete specifications}
It is well known that test-based equivalence is prone to issues of incomplete coverage. 
Accordingly, our test-based validation (with randomly sampled tests) also has some chance of missing bugs in translation.
We perform coverage analysis for our test suite and report the numbers in \secref{sec:experiments}.
For large (multi-file, multi-function) codebases, test-based/fuzzing-based validation remains the primary scalable testing strategy.
Future work can improve testing by using better sampling/fuzzing heuristics to improve coverage.
%

\Paragraph{Test-based Equivalence: Overly strict inner equivalence}
While test-based validation of the final translation can lead to incomplete coverage, enforcing equivalence at intermediate functions results in the opposite issue: it may be unnecessarily strict.
Relaxing this may result in simpler, more interpretable/performant code.
However, determining which equivalence checks to relax (and how) is challenging and requires knowledge of future (downstream) context.
Our current dependency-ordered greedy translation approach provides limited flexibility for such relaxations.
%

\paragraph{Example.} A particularly interesting instance is the \rustc{zopfli_append_data} function.
The \clanguage{} code calls \cc{realloc} to double the allocated memory of arrays in case the size runs out while appending elements (mimicking vectors) and stores the current size of the array separately.
In \rustlang{}, \rustc{Vec} is a natural choice for mapping the \clanguage{} array, which internally performs the doubling logic. 
However, to maintain equivalence between the \clanguage{} and \rustlang{} code, the \rustlang{} code also needs to artificially double the array size and use a separate variable to store the \textit{size} of the vector.
Notably, this leads to non-idiomatic \rustlang{} code, and the \llm{} confuses \rustc{Vec.length} as the size of the vector instead of using the secondary size stored in a variable. 

\Paragraph{Cascading post-generation repairs}
Recall that a set of equivalence tests guides our greedy dependency-ordered translation.
Once we have generated the complete Rust codebase (post-generation), we validate the entire translation by performing equivalence tests from the top-level codebase entry point.
However, some of these validation tests may still fail. We discuss such a case in our experimentation section (\secref{sec:experiments}).
While we can localize the error using our \eqtester infrastructure, repairs to the erroneous function may cascade to its dependent functions.
In such cases, we want to perform \textit{minimal} edits to the translated codebase so that the validation tests pass.
While we manually strategize this in our experiments, future work can develop more systematic techniques for \llm{}-guided repair, viewing this as an agentic \llm{} system with edit primitives/actions.

\Paragraph{Type Punning and \cc{void*}}
C provides full memory control using manually constructed addresses. Interpreting programs written in this way is difficult and forms an active research area (e.g. software reverse engineering survey \cite{nelson2005survey}).
Even when a program does not use raw memory accesses, \cc{void*} pointers can obfuscate information about the underlying object.
Our approach, particularly our dynamic analyses (e.g., 
\secref{subsec:dynamicspec}, \hyperref[para:types]{type analysis}) relies on information being available during compilation.
Thus, type punning and rampant usage of \cc{void*} can cause imprecise/incomplete analysis and lead to incorrect type inference.
While this forms a limitation, we do not see a straightforward solution.
We believe that future work that aims to resolve this would need to leverage reverse engineering research/powerful (and likely costly) program analyses.

\Paragraph{Cyclic Structs and Multi-threading}
Our current approach does not support cyclic C structs and multi-threading. 
Cyclic structs require careful handling in Rust with \rustc{Weak} references to break cycles and avoid memory leaks.
Similarly, multi-threading requires careful handling of shared mutable state and synchronization primitives using \rustc{Arc, Mutex} primitives.
While \llms{} can theoretically adapt to these cases, implementing the necessary program analysis to enable specification mining and testing for these cases will require further work.

\Paragraph{Argument Translation API}
We design an argument translation API to translate C arguments to Rust arguments (\secref{subsubsec:argumenttranslation}).
The API provides higher order primitives, which allow translating primitive types and multi-dimension vectors (of arbitrary types) while taking care of aliasing (using \rustc{Rc<RefCell<RT>>}).
We identify a simplicity-expressiveness trade-off in the API design.
We can provide powerful and general high-level primitives that can handle a wide range of operations but may be difficult to use (requiring lambda function arguments).
Alternatively, we can provide simple and easy-to-use primitives that can handle only a limited set of operations.
Here, we identify a simple API design that handles the most common cases and allows \llms{} to generate arbitrary \rustlang{} code for complex cases.
However, future work can explore this tradeoff further, improving the API design by making it easier for \llms{} to use and more powerful.

\Paragraph{Efficiency of Translated Program}
Our current approach does not focus on the efficiency of the translated program.
Indeed, our translated \zopfli{} program is up to $3.67\times$ slower than the original C program.
We attempt to identify the source of the slow-down by comparing the programs statically and using corresponding flame-graphs but do not observe meaningful insights to translate faster. 
Going forward, we envision
two approaches to improve the efficiency of the translated program.
One approach is to translate the program and then use profiling to identify and optimize the slow parts of the program.
Another approach is to ensure that the translated program is efficient from the start.
We aim to explore these directions going forward.

\subsection{Threats to validity}
\Paragraph{External Validity}
We demonstrate the efficacy of our translation approach by successfully translating \zopfli{} and \urlparser{}.
This can cause over-fitting in the prompts and impact design decisions in our implementation.  
First note that \zopfli{} is over $3000$ LoC and contains diverse code elements encompassing linked lists, iterators, function pointers, \cc{void*} elements.
Secondly, the ~\sysname{} approach is quite generic and only makes moderate assumptions about the \clanguage{} programs.
Nevertheless, as we extend \sysname{}, we will evaluate it on a broader suite of programs going forward.

\Paragraph{Test-based equivalence}
We assert the validity of our translation for \zopfli{} using test cases. 
However, test-based equivalence is known to suffer from incomplete coverage.
Here, we generated over $\numzopflitestseval$ tests to achieve high coverage to strengthen our claims.

\subsection{Conclusion}
In this work, we presented \sysname{}, our approach that synergistically combines program analysis (specifically dynamic analysis and testing) with \llm{} sampling-driven-search approach to translate \clanguage{} programs to \rustlang{}. 
Particularly, we use the program dependency graph to decompose translation into individual translation units and perform dual code and test translation to generate valid \rustlang{} code.
We demonstrate the efficacy of our approach by correctly translating \zopfli{}, a data compression \clanguage{} program with over $3000$ LoC.
Our approach outlines a way to \textit{naturally} perform test-driven development and is scalable.
In particular, improving the empowering our approach with better dynamic analysis and stronger \llm{} inference time schedules will allow us to translate more and more complex programs.

\Paragraph{Acknowledgement}
This work is generously supported by the OpenPhilanthropy R2E grant. 
Naman and Manish are supported by National Science Foundation grant CCF-1900968 and SKY Lab industrial sponsors
and affiliates Astronomer, Google, IBM, Intel, Lacework,
Microsoft, Mohamed Bin Zayed University of Artificial
Intelligence, Nexla, Samsung SDS, Uber, and VMware.
Adwait is supported by the Intel Scalable Assurance program, DARPA contract FA8750-20-C0156, and NSF grant 1837132.
Finally, we thank Alex Gu, Wen-Ding Li, Hao Wang, Sida Wang, Federico Cassano, and Pei-Wei Chen for helpful comments and feedback on the work.

\ifx\FORMAT\ACM

\bibliographystyle{ACM-Reference-Format}
\bibliography{refs}

\else

\bibliographystyle{IEEEtran}
\bibliography{refs}
\fi
\newpage
\appendix
\section{Prompts}
In the following, we present prompts for different modules of our approach (please see next page).
\begin{figure}[h!]
\centering
\begin{lstlisting}[style=PStyle, xleftmargin=0cm, basicstyle=\fontencoding{T1}\fontfamily{lmtt}\scriptsize]
You are given the following C code:
```c
{c_decls}
{c_stmts}
```

You have already translated the following C code to Rust code:
```rust
{rust_stmts}
```

Now you will translate the following C code to Rust code:
```c
{c_symbol}
{dynamic analysis annotations}
```

You will translate the C code to SAFE Rust and return it as the first code block. Follow these guidelines:
  1. You should only translate the given C code element to Rust code.
  2. Do NOT modify or rewrite any existing Rust code provided above. Your translation should be designed to be appended to the existing Rust code. Write the Rust code only for the given element so that it can be appended to the existing Rust code. Do NOT rewrite existing rust code since it will cause compile errors.
  3. Precisely follow the program logic, function signatures, types and returns in the C program. Do NOT make unnecessary modifications or simplifications to the SAFE Rust code. MINIMIZE the differences between the C and Rust code and maintain correspondence between the function signatures as much as possible. Since, we are aiming for a direct translation of a large codebase it would be ideal to not make large refactoring changes.
  4. Do NOT leave placeholder, empty, or incomplete code. Do NOT skip long code blocks. Ensure that the Rust code is complete and correct.
  5. Create struct methods only when absolutely necessary. Prefer implementing functions over structs. If struct methods are needed, write a new `impl` block enclosing only that method. Do NOT copy reimplement or copy struct definition or existing struct methods. Do not implement additional traits on structs.
  6. Translate C arrays to Rust vectors whenever possible. Use Rust arrays only when the array size is fixed and known at compile time. For example, 
     - Translate `int arr[10]` to `[i32; 10]` since size is fixed and known at compile time.
     - Translate `int* arr` to `Vec<i32>` since size is not fixed and known at runtime.
     - Translate `int *(arr)[10]` to `[Vec<i32>; 10]` and `int (*arr)[10]` to `Vec<[i32; 10]>`.
     - Do NOT use `Option` inside the vector. `Vec<Option<T>>` can be simulated using a 0 length `Vec<T>`.
     - In case the length of the vector along a dimension is always supposed to be 1, you can 'squeeze' the dimension and construct a smaller dimensional vector. However, in case it is not clear, always use the full dimension.
  7. Translate function pointers to `Fn` or `FnMut` closures.
  8. Translate void pointers to either concrete types (if clear from context) or use `dyn std::any::Any`. Do NOT use `*mut c_void`. The `Any` can later be resolved using `downcast_ref` to downcast to the concrete type when C code casts the void pointer to a specific type.
  9. Use the existing translated Rust structs and functions provided above. Additionally, `NodePair` struct has been added to Rust code to simplify the translation of `lists` which can be translated to Vec<NodePair>.
  10. Think step by step and reason about the translated rust code in natural language. Add detailed comments sketching out logic of the program in the Rust code. Maintain equivalence between the C and the Rust programs, logic, and signature as much as possible.
  11. Add comments both at the start of the rust code block (documenting a function) as well as inlined in the code block. Always generate public rust code (annotated with `pub`).
  12. Follow the logic of the C code and do NOT make any optimizations or unnecessary changes. Maintain 1-1 correspondence between the C and Rust code blocks as much as possible. If the C code maintains a variable for length of the array, maintain the same in Rust. If the C code indexes into an array, write the Rust code performing indexing similarly. MAINTAIN the nullability and aliasing across C and the Rust code as provided above in comments. More generally, perform the SIMPLEST AND THE MOST DIRECT translation possible ENSURING CORRECT BEHAVIOUR with SAFE Rust.

Rust Reminder: You cannot have a mutable and immutable reference to the same data at the same time. If you need to mutate data, you should use `Rc` `RefCell` to provide interior mutability and shared ownership. Alternatively, you can refactor the code to avoid the need for shared mutability.
\end{lstlisting}
\caption{Rust Code Generation Prompt}
\end{figure}
\begin{figure}[h!]
\centering
\begin{lstlisting}[style=PStyle, xleftmargin=0cm, basicstyle=\fontencoding{T1}\fontfamily{lmtt}\scriptsize]
Following is the C code you are provided:
```c
{c_decls}
{c_stmts}
```

Following is the translated Rust code
```rust
{final_rust_code}
```

We will use functional equivalence to check the correctness of the translation of `{function_name}` function using FFI. Thus, given a C function `{function_name}` and C arguments, we will first convert the C arguments to Rust arguments and then call the Rust function with these arguments. This is useful to compare the behavior of the Rust function with the expected behavior of the C function. For now, you will only translate the arguments.

To aid with translation, you are provided with the following utilities:
[TRANSLATION_API_DOCS_AND_EXAMPLES]

Now, you will implement the `translate_args` function for `{function_name}`. Reminder: `{function_name}` signature:
```c
{function_declaration}
```
You will enclose your answer in a markdown code block. Follow the signature below. Feel free to modify the mutability of the returned values as needed.
```rust
{translate_signature}
```

Follow these instructions
1. **Scalar arguments**. You can use the `translate_singleton` and `translate_singleton_and_peel` functions to translate scalar arguments. These functions return `Option<T>` and `T` respectively. Additionally, you can use `expect` to remove the wrapping `Option` as required.
2. **N-dimensional arrays**. You can use the `translate_nd_vec` and `translate_nd_vec_and_peel` functions to translate n-dimensional arrays. These functions return `Vec<Vec<...Rc<RefCell<T>>...>>` and `Vec<Vec<...T...>>` respectively. In cases where a dimension is 1, you can "squeeze" the dimension by indexing along that dimension. For example, `translate_2d_vec_and_peel` returns `Vec<Vec<T>>` and you can get the first row using `[0]` indexing.
3. **Types**. These functions work across both primitive types and structs. Therefore, you do NOT need to implement custom translations for arbitrary structions. However, you might need to massage the output of the translation functions to match the expected Rust types.
4. **Allowed Types**. The type T can be primitive types (`i8`, `i16`, `i32`, `i64`, `u8`, `u16`, `u32`, `u64`, `f32`, `f64`) or structs (... enumerate ...). You can construct singleton instances of these types using the `translate_singleton` and vectors of these types using the `translate_nd_vec` functions. Note that `NodePair` can be used to directly implement argument translation for `lists` (using `translate_1d_vec`). Thus Node *(*lists)[2] can be translated to `Vec<NodePair>` via `translate_1d_vec_and_peel(c_lists)`.
5. **C Structs**. Note that the structs in the C code are already imported in Rust using bindgen + FFI. They have `C` prefix attached to them. For example, `ZopfliOptions` in C is imported as `CZopfliOptions` in Rust, `Node` in C is imported as `CNode` in Rust. You can use these structs directly in the translation functions. Note that the Rust structs are named without the `C` prefix (e.g., `ZopfliOptions`, `Node`).
6. **Aliasing**. The translation functions use a symbol table internally to handle aliasing. The aliasing is captured by the `Rc<RefCell<T>>` type. Therefore, you do NOT need to handle aliasing explicitly. In cases aliasing is not present, you can use the `*_peel` functions to get the inner value directly.
7. **Return By Value**. Do NOT return `&mut T` from the translation functions since the variable will not be live after the function is returned. Instead, return the value directly or use `Rc<RefCell<T>>` for shared ownership and interior mutability.
8. **Function Pointers**. Function pointers can be resolved to function names. You can use a match statement to translate to the appropriate function name.
9. **Void Pointers**. Void pointers either can be resolved to a specific type (evident from the context) or can be translated to `dyn std::any::Any`. In the latter case, you can use the function name to determine the type of the void pointer to construct rust object and then upcast it to `dyn Any`.

\end{lstlisting}
\caption{TranslateArgs Prompt}
\end{figure}
\begin{figure}[h!]
\centering
\begin{lstlisting}[style=PStyle, xleftmargin=0cm, basicstyle=\fontencoding{T1}\fontfamily{lmtt}\scriptsize]
Following is the C code you are provided:

```c
{c_decls}
{c_stmts}
```

Following is the translated Rust code

```rust
{final_rust_code}
```

We are going to use functional equivalence to check the correctness of the translation of `{c_fn_name}` function.

Reminder of the C function signature:
```c
{c_fn_decl}
```
Instructions:
1. In the equivalence test function, to translate C arguments to Rust arguments call `translate_args_{c_fn_name}` which you can assume already exists in the same file.
2. Compare equivalence of BOTH the return values and the values of the arguments after the function call as appropriate.
3. If needed use `translate_args_{c_fn_name}` to convert C function results to Rust **after** the C function call to be able to compare with the Rust function's results.
4. If any results (argument and returned values) are of type Box<dyn Any>, you should downcast them to the expected type using the `downcast_ref::<T>` method since \'==\' operator does not work on trait objects. However, you should ensure that the downcast type is properly handled and downcast runs successfully.

Here is the translate_args function:
```rust
{translate_args_fn}
```

Your equivalence test function `eqtest_{c_fn_name}` will be called with the following harness which already handles loading C arguments for the function. 

```rust
{harness}
```

Assuming the above harness, now you will only implement the equivalence testing function which should:
   - load the C arguments from the given file using the following statement:
          ```let ({c_pre_args}) = load_pre_args_from_json(file_name);```
   - use the `translate_args_{c_fn_name}` function to convert C arguments to Rust types.
   - load the expected values of the C arguments from the given file using the following statement:
          ```let ({c_post_args}) = load_post_args_from_json(file_name);```
    - use the `translate_args_{{c_fn_name}}` function to convert expected C arguments to Rust types. Use the appropriate types for the expected values as per the translate_args function signature.
   - call the `{c_fn_name}` C function with C arguments assuming the FFI bindings are declared in the bindings crate. Use this only to get the return values.
   - call the `{rust_fn_name}` rust function with translated Rust arguments.
   - compare the results for functional equivalence using assertions with the `assert_eq!` macro. You must use assertion messages to show what is being compared.
   - return true if all assertions pass. Otherwise, let the code panic naturally.

Only return the equivalence testing function and nothing else. You will enclose your answer in a markdown code block.
```rust
{eqtest_signature}
```

Ensure that you correctly test the function and check the outputs match the C outputs (if the function returns a value). Additionally, if the function modifies the inputs, perform equivalence checks on the modified (mutable) inputs appropirately. You DO NOT need to do equivalence on the immutable inputs.

\end{lstlisting}
\caption{Rust Equivalence Test Generation Prompt}
\end{figure}

\section{Example}
In the following, we present an example of translating the \cc{GetCostModelMinCost} function in \zopfli{}. We show the following artefacts from our pipeline:
\begin{enumerate}
    \item \textbf{Source C function} \cc{GetCostModelMinCost}: (Figure \ref{fig:cost-model-C})
    \item \textbf{Candidate Rust translation} \rustc{get_cost_model_min_cost}: (Figure \ref{fig:cost-model-Rust})
    \item \textbf{Argument translator}: (Figure \ref{fig:cost-model-translateargs})
    \item \textbf{Equivalence test}: (Figure \ref{fig:cost-model-eqtest})
\end{enumerate}

\begin{figure}
\begin{lstlisting}[style=CStyle, xleftmargin=0cm, basicstyle=\fontencoding{T1}\fontfamily{lmtt}\scriptsize]
/*
Finds the minimum possible cost this cost model can return for valid length and
distance symbols.
*/
static double GetCostModelMinCost(CostModelFun *costmodel, void *costcontext)
{
  double mincost;
  int bestlength = 0; /* length that has lowest cost in the cost model */
  int bestdist = 0;   /* distance that has lowest cost in the cost model */
  int i;
  /*
  Table of distances that have a different distance symbol in the deflate
  specification. Each value is the first distance that has a new symbol. Only
  different symbols affect the cost model so only these need to be checked.
  See RFC 1951 section 3.2.5. Compressed blocks (length and distance codes).
  */
  static const int dsymbols[30] = {
      1, 2, 3, 4, 5, 7, 9, 13, 17, 25, 33, 49, 65, 97, 129, 193, 257, 385, 513,
      769, 1025, 1537, 2049, 3073, 4097, 6145, 8193, 12289, 16385, 24577};

  mincost = ZOPFLI_LARGE_FLOAT;
  for (i = 3; i < 259; i++)
  {
    double c = costmodel(i, 1, costcontext);
    if (c < mincost)
    {
      bestlength = i;
      mincost = c;
    }
  }

  mincost = ZOPFLI_LARGE_FLOAT;
  for (i = 0; i < 30; i++)
  {
    double c = costmodel(3, dsymbols[i], costcontext);
    if (c < mincost)
    {
      bestdist = dsymbols[i];
      mincost = c;
    }
  }

  return costmodel(bestlength, bestdist, costcontext);
}
\end{lstlisting}
\centering
\caption{The \cc{GetCostModelMinCost} function from the \zopfli{} codebase. The function's complexity stems from using indirect function calls through the \cc{costmodel} function pointer coupled with a \cc{void*} context parameter. Particularly hard is to interpret (purely statically) that the function pointer can point to several cost functions.}
\label{fig:cost-model-C}
\end{figure}


\begin{figure}
\begin{lstlisting}[style=RStyle, xleftmargin=0cm, basicstyle=\fontencoding{T1}\fontfamily{lmtt}\scriptsize]
/// Finds the minimum possible cost this cost model can return for valid length and
/// distance symbols.
///
/// # Arguments
/// * `costmodel` - A reference to the cost model function or closure.
/// * `costcontext` - A reference to any context-specific data required by the cost model.
///
/// # Returns
/// Returns the minimum possible cost as an `f64`.
///
/// # Notes
/// - The function uses `ZOPFLI_LARGE_FLOAT` as an initial high cost value.
/// - The best length and distance are found separately and then combined to compute the final cost.
pub fn get_cost_model_min_cost(
    costmodel: &dyn Fn(u32, u32, &dyn std::any::Any) -> f64,
    costcontext: &dyn std::any::Any,
) -> f64 {
    // Initialize variables for minimum cost and best length/distance.
    let mut min_cost: f64;
    let mut best_length: u32 = 0; // Length that has the lowest cost in the cost model.
    let mut best_dist: u32 = 0;   // Distance that has the lowest cost in the cost model.

    /*
    Table of distances that have a different distance symbol in the DEFLATE
    specification. Each value is the first distance that has a new symbol. Only
    different symbols affect the cost model, so only these need to be checked.
    See RFC 1951 section 3.2.5: Compressed blocks (length and distance codes).
    */
    let dsymbols: [u32; 30] = [
        1, 2, 3, 4, 5, 7, 9, 13, 17, 25,
        33, 49, 65, 97, 129, 193, 257, 385,
        513, 769, 1025, 1537, 2049, 3073,
        4097, 6145, 8193, 12289, 16385, 24577,
    ];

    // Find the length with the lowest cost.
    min_cost = ZOPFLI_LARGE_FLOAT;
    for i in 3..259 {
        let c = costmodel(i as u32, 1, costcontext);
        if c < min_cost {
            best_length = i as u32;
            min_cost = c;
        }
    }

    // Find the distance with the lowest cost.
    min_cost = ZOPFLI_LARGE_FLOAT;
    for &dist in dsymbols.iter() {
        let c = costmodel(3, dist, costcontext);
        if c < min_cost {
            best_dist = dist;
            min_cost = c;
        }
    }

    // Return the minimum possible cost using the best length and distance found.
    costmodel(best_length, best_dist, costcontext)
}
\end{lstlisting}
\centering
\caption{A candidate Rust translation generated for the C function \cc{GetCostModelMinCost}}
\label{fig:cost-model-Rust}
\end{figure}


\begin{figure}
\begin{lstlisting}[style=RStyle, xleftmargin=0cm, basicstyle=\fontencoding{T1}\fontfamily{lmtt}\scriptsize]
pub unsafe fn translate_args_GetCostModelMinCost(
    costmodel_ptr: Option<
        unsafe extern "C" fn(
            litlen: ::std::os::raw::c_uint,
            dist: ::std::os::raw::c_uint,
            context: *mut ::std::os::raw::c_void,
        ) -> f64,
    >,
    costcontext: *mut ::std::os::raw::c_void,
) -> (
    Box<dyn Fn(u32, u32, &dyn std::any::Any) -> f64>,
    Box<dyn std::any::Any>,
) {
    rs_symtab_clear();

    // Get the function name from the function pointer
    let fn_ptr = costmodel_ptr.expect("Function pointer is null");
    let fn_name = get_fn_name_from_fn_ptr(fn_ptr as *mut c_void)
        .expect("Failed to get function name from function pointer");

    if fn_name == "GetCostFixed" {
        // For GetCostFixed, no context is needed
        // Create a closure that matches the expected signature
        let costmodel = Box::new(|litlen: u32, dist: u32, _context: &dyn Any| -> f64 {
            GetCostFixed(litlen, dist, std::ptr::null_mut())
        }) as Box<dyn Fn(u32, u32, &dyn std::any::Any) -> f64>;

        // The context is not used, so we can use an empty tuple
        let costcontext = Box::new(()) as Box<dyn Any>;

        (costmodel, costcontext)
    } else if fn_name == "GetCostStat" {
        // For GetCostStat, the context is a SymbolStats structure
        // Translate the context from C to Rust
        let rust_context = translate_singleton_and_peel::<SymbolStats>(costcontext)
            .expect("Failed to translate SymbolStats context");

        // Create a closure that captures the context and matches the expected signature
        let costmodel = Box::new(move |litlen: u32, dist: u32, context: &dyn Any| -> f64 {
            // Downcast the context to SymbolStats
            let stats = context
                .downcast_ref::<SymbolStats>()
                .expect("Failed to downcast context to SymbolStats");
            zopfli_get_cost_stat(litlen as i32, dist as i32, stats)
        }) as Box<dyn Fn(u32, u32, &dyn std::any::Any) -> f64>;

        // Box the context so it can be used as a dynamic Any type
        let costcontext = Box::new(rust_context) as Box<dyn Any>;

        (costmodel, costcontext)
    } else {
        panic!("Unknown function name: {}", fn_name);
    }
}
\end{lstlisting}
\centering
\caption{The \llm{} generated \translateargs{} function for the \cc{GetCostModelMinCost}, handling the translation of C function pointers to Rust closures, supporting multiple variants the pointer may point to: \cc{GetCostFixed} and \cc{GetCostStat}. The \translateargs{} is aligned with the candidate translation in Fig. \ref{fig:cost-model-Rust}}.
\label{fig:cost-model-translateargs}
\end{figure}


\begin{figure}
\begin{lstlisting}[style=RStyle, xleftmargin=0cm, basicstyle=\fontencoding{T1}\fontfamily{lmtt}\scriptsize]
pub unsafe fn eqtest_GetCostModelMinCost(filename: &str) -> bool {
    let (c_pre_arg0, c_pre_arg1) = load_pre_args_from_json(filename);

    // Translate the C arguments to Rust arguments using the provided translate function
    let (rust_costmodel, rust_costcontext) =
        translate_args_GetCostModelMinCost(c_pre_arg0, c_pre_arg1);

    // Call the C function with the original C arguments
    let c_result = bindings::GetCostModelMinCost(c_pre_arg0, c_pre_arg1);

    // Call the Rust function with the translated Rust arguments
    let rust_result = get_cost_model_min_cost(&*rust_costmodel, &*rust_costcontext);

    // Compare the return values of C and Rust functions, allowing for possible floating-point differences
    let epsilon = 1e-6;
    assert!(
            (c_result - rust_result).abs() < epsilon,
        "C and Rust functions returned different values: C={} vs Rust={}",
        c_result,
        rust_result
    );

    // Since the arguments are not modified by the function, no need to compare them after the call

    // All assertions passed
    true
}
\end{lstlisting}
\centering
\caption{The \llm{} generated equivalence test for the C function \cc{GetCostModelMinCost} (Fig. \ref{fig:cost-model-C}) and candidate translation \rustc{get_cost_model_min_cost} (Fig. \ref{fig:cost-model-Rust}). Note the use of the previously generated \translateargs{} (Fig. \ref{fig:cost-model-translateargs}).
Also, the model-generated test uses assertions with tolerance for floating-point precision.
}
\label{fig:cost-model-eqtest}
\end{figure}

\end{document}